\documentclass[11pt]{elsarticle}

\usepackage{float}
\usepackage{subfigure}
\usepackage{lineno}
\usepackage{amsmath,amsfonts,amssymb,bm}
\usepackage{pslatex}
\usepackage{graphicx,color}
\usepackage{xcolor}
\usepackage{fancyhdr}
\usepackage[symbol]{footmisc}
\usepackage[colorlinks=true,linkcolor=blue]{hyperref}%
\usepackage{gensymb}
\usepackage[all]{nowidow}
\newcommand{\n}[1] {\mbox{\boldmath{$#1$}}} 
\newcommand{\wh}{\widehat}
\def\E{\mathbb E}
\DeclareMathOperator{\cov}{cov}
\DeclareMathOperator{\Tr}{tr}
\DeclareMathOperator{\diag}{diag}
\def\N{\mathcal N}
\def\R{\mathbb R}
\def\R{{\small\textsf{R}}}

\usepackage{amsthm}
\theoremstyle{definition}
\newtheorem{definition}{Definition}[]
\newtheorem{theorem}{Theorem}[]
\usepackage{natbib}

\journal{Spatial Statistics}

\biboptions{authoryear}

\begin{document}

\begin{frontmatter}

\title{A Spatial Concordance Correlation Coefficient with an Application to Image Analysis}

\author{Ronny Vallejos$^{1}$\footnote[1]{Corresponding author: R. Vallejos, \texttt{ronny.vallejos@usm.cl}}, Javier P\'erez$^{2},$ Aaron M. Ellison$^3$ and Andrew D. Richardson$^4$}
\address{$^{1,2}$ Departamento de Matem\'atica, Universidad T\'ecnica Federico Santa Mar\'ia, Avenida Espa\~na 1680, Valpara\'iso, Chile\\
  $^3$ Harvard Forest, Harvard University, Petersham, Massachusetts, USA \\ 
$^4$ School of Informatics, Computing and Cyber Systems, Northern Arizona University, USA and Center for Ecosystem Science and Society, Northern Arizona University, USA \\
}




\begin{abstract}
In this work we define a spatial concordance coefficient for second-order stationary processes. This problem has been widely addressed in a non-spatial context, but here we consider a coefficient that for a fixed spatial lag  allows one to compare two spatial sequences along a 45$^{\circ}$ line.  The proposed coefficient was explored for the bivariate Mat\'ern and Wendland covariance functions. The asymptotic normality of a sample version of the spatial concordance coefficient for an increasing domain sampling framework was established for the Wendland covariance function. To work with large digital images, we developed a local approach for estimating the concordance that uses local spatial models on non-overlapping windows. Monte Carlo simulations were used to gain additional insights into the asymptotic properties for finite sample sizes. As an illustrative example, we applied this methodology to two similar images of a deciduous forest canopy. The images were recorded with different cameras but similar fields-of-view and within minutes of each other. Our analysis showed that the local approach helped to explain a percentage of the non-spatial concordance and to provided additional information about its decay as a function of the spatial lag.
\end{abstract}


\end{frontmatter}


\section{Introduction}
In recent decades, concordance correlation coefficients have been developed in a variety of different contexts. For instance, in assay or instrument validation processes, the reproducibility of the measurements among trials or laboratories is of interest. When a new instrument is developed, it may be relevant to evaluate whether its performance is concordant with other, existing ones, or its results accord with a ``gold standard.'' There are also situations in which one is interested in comparing two methods without a designated gold standard or target values \citep{Lin:2002}. In the literature, this latter type of concordance has been tackled  from different perspectives \citep{Barnhart:2007}. \cite{Cohen:1968} discussed this problem in the context of categorical data. \cite{Schall:1996} and \cite{Lin:2000} performed similar studies in the context of bioequivalence.

One way to approach the concordance problem for continuous measurements is to construct a scaled summary index that can take on values between $-1$ and 1, analogous to a correlation coefficient. Using this approach, \cite{Lin:1989} suggested a concordance correlation coefficient (CCC) that evaluates the agreement between two continuous variables by measuring their joint deviation from a 45$^\circ$ line through the origin. There have been some extensions of this CCC that use several measuring instruments and techniques to evaluate the agreement between two instruments; these efforts have led to interesting graphical tools \citep{Hiriote:2011,Stevens:2017}. In the context of goodness of fit, \cite{Vonesh:1996} proposed a modified Lin's CCC for choosing models that have a better agreement between observed and the predicted values. Recently \cite{Stevens:2017} and \cite{Choudhary:2017} developed the probability of
agreement, and \cite{Leal:2019} studied the local influence of the CCC and the probability of agreement considering both first- and second-order measures under the case-weight perturbation scheme.  \cite{Atkinson:1997} critiqued the CCC because any correlation coefficient is highly dependent on the measurement range. In general, therefore, CCC is used only when measuring ranges are comparable or when methods are on the same scale.

In this paper, we suggest an approach to assessing the agreement between two continuous responses when the observations of both variables have been georeferenced in space. We define a spatial CCC (SCCC) as a generalization of Lin's (\citeyear{Lin:1989}) coefficient that measures the agreement between two spatial variables. For a fixed lag, our SCCC shares the same properties as the original CCC. For an increasing sampling scheme, we establish the asymptotic normality of the sample SCCC for a bivariate Gaussian process with a Wendland covariance function. To improve the behavior of the coefficient, we developed a local approach for estimating it that uses local spatial models on non-overlapping windows. This approach constitutes a new way of thinking about concordance that has not been considered previously, especially for large digital images. Our approach also captures the decay of the SCCC as a function of the norm of the spatial lag. Monte Carlo simulations and numerical experiments with real datasets accompany the exposition of the methodological aspects. An image-analysis example is worked in detail to illustrate the fitting of a local SCCCs. We conclude with a summary of the main findings and an outline of problems to be tackled in future research.  

\section{Preliminaries and Notation}
Assume that $X$ and $Y$ are two continuous random variables such that the joint distribution of $X$ and $Y$ has finite second moments with means  $\mu_{X}$ and $\mu_{Y}$, variances $\sigma^2_X$ and $\sigma_Y^2$, and covariance $\sigma_{YX}$. The mean squared deviation of $D=Y-X$ is
$$\text{MSD}=\epsilon^2=\mathbb{E}[D^2]=\mathbb{E}[(Y-X)^2].$$
It is straightforward to see that $\epsilon^2=(\mu_X-\mu_Y)^2+\sigma_Y^2+\sigma_X^2-2\sigma_{YX}$ and the sample counterpart satisfies $e^2=(\overline{y}-\overline{x})^2+s_Y^2+s_X^2-2s_{XY}.$ Using this framework, \cite{Lin:1989} defined a CCC as:
\begin{equation}\label{eq:Lin}
\rho_c=1-\frac{\epsilon^2}{\epsilon^2|\rho=0}=\frac{2\sigma_{YX}}{\sigma_Y^2+\sigma_X^2+(\mu_Y^2-\mu_X^2)^2}.
\end{equation}
The CCC satisfies the following properties:
\begin{itemize}
\item[1.] $\rho_c=\alpha \cdot \rho,$ where $\alpha=\frac{2}{w+1/w+v^2}$ and $w=\frac{\sigma_Y}{\sigma_X}.$
\vspace{-2mm}
\item[2.] $|\rho_c|\leq 1.$
\vspace{-2mm}
\item[3.] $\rho_c=0$ if and only if $\rho=0.$
\vspace{-2mm}
\item[4.] $\rho_c=\rho$ if and only if $\sigma_Y=\sigma_X$ and $\mu_Y=\mu_X$.
\end{itemize}

The sample estimate of $\rho_c$ is given as
$$\widehat{\rho}_c=\frac{2 s_{YX}}{s_Y^2+s_X^2+(\overline{y}-\overline{x})^2}.$$
The inference for this coefficient was addressed via Fisher's transformation. \cite{Lin:1989} proved that $$Z=\frac{1}{2}\left(\frac{1+\widehat{\rho}_c}{1-\widehat{\rho}_c}\right)\stackrel{\mathcal{D}}{\longrightarrow} \mathcal{N}(\psi,\sigma_Z^2), \ \text{as} \ n\rightarrow \infty,$$
where
$$\psi=\tanh^{-1}(\rho_c)=\frac{1}{2}\left(\frac{1+\rho_c}{1-\rho_c}\right),$$ 
$$\sigma_Z^2=\frac{1}{n-2}\left[\frac{(1-\rho^2)\rho_c^2}{(1-\rho_c^2)\rho^2}+\frac{2v^2(1-\rho_c)\rho_c^3}{(1-\rho_c^2)^2\rho}+\frac{v^4\rho_c^4}{2(1-\rho_c^2)^2\rho^2}\right],$$ and
$$v^2=\frac{(\mu_Y-\mu_X)^2}{\sigma_Y \sigma_X}.$$

As a consequence of the asymptotic normality of the sample CCC, an approximate hypothesis testing problem of the form $$\text{H}_0:\rho_c=\rho_0 ~ \text{\textit{versus}} ~ \text{H}_1:\rho_c \neq \rho_0$$ for a fixed $\rho_0$ can be constructed. Alternatively, an approximate confidence interval of the form 
$$\widehat{\rho}_c\pm z_{\alpha/2}\sqrt{\sigma_Z^2}$$ can be used, where $z_{\alpha/2}$ is the upper quantile of order $\alpha/2$ of the standard normal distribution. Applications and extensions of Lin's coefficient can be found in \cite{Lin:2012}, among others.


\section{A Spatial Concordance Coefficient and its Properties}\label{sec:Spatial_cor}

We start by extending Lin's CCC for bivariate second-order spatial processes for a fixed lag in space.
\begin{definition}
Let $(X(\bm s), Y(\bm s))^{\top}$ be a bivariate second-order stationary random field  with  $\bm s \in \mathbb{R}^2$, mean $(\mu_1,\mu_2)^{\top}$, and covariance function
$$C(\bm h)=\left(\begin{matrix} C_X(\bm h) & C_{XY}(\bm h)\\
C_{YX}(\bm h) & C_{Y}(\bm h)
\end{matrix} \right).$$
Then the SCCC is defined as 
\begin{align} \label{eq:spatial_cor}
\rho^c(\bm h)&=\frac{\mathbb{E}[(Y(\bm s+\bm h)-X(\bm s))^2]}{\mathbb{E}[(Y(\bm s+\bm h)-X(\bm s))^2|C_{XY}(\bm 0)=0]} \notag\\
&=\frac{2C_{YX}(\bm h)}{C_{X}(\bm 0)+C_{Y}(\bm 0)+(\mu_1-\mu_2)^2}.
\end{align}

\end{definition}
Some straightforward properties of this SCCC are:
\begin{itemize}
\item[1.] $\rho^c(\bm h)=\eta \cdot \rho_{YX}(\bm h),$ where
 $\eta=\frac{2 \sqrt{C_X(\bm 0)C_Y(\bm 0)}}{C_X(\bm 0)+C_Y(\bm 0)+(\mu_1-\mu_2)^2}.$
\vspace{-2mm}
\item[2.] $|\rho^c(\bm h)|\leq 1.$
\vspace{-2mm}
\item[3.] $\rho^c(\bm h)=0$ iff $\rho_{YX}(\bm h)=0.$
\vspace{-2mm}
\item[4.] $\rho^c(\bm h)= \rho_{YX}(\bm h)$ iff $\mu_1=\mu_2$ and $C_X(\bm 0)=C_Y(\bm 0).$
\vspace{-2mm} 
\item[5.] For a bivarite Mat\'ern covariance function defined as \citep{Gneiting:2010} 
\begin{align}
C_X(\bm h)&=\sigma_1^2 M(\bm h,\nu_1,a_1), \label{eq:mat1}\\ C_Y(\bm h)&=\sigma_2^2 M(\bm h,\nu_2,a_2),\label{eq:mat2}\\ \mu_1&=\mu_2, \notag\\
C_{YX}(\bm h, \nu_{12},a_{12})&=\rho_{12}\sigma_1 \sigma_2M(\bm h,\nu_{12},a_{12}),\label{eq:mat3}
\end{align}
where $M(\bm h,\nu,a)=(a||\bm h||)^{\nu}K_{\nu}(a||\bm h||)$, $K_{\nu}(\cdot)$ is a modified Bessel function of the second kind, and $\rho_{12}=\text{cor}[X(\bm s_i), Y(\bm s_j)]$, it follows that
\begin{align*}
\rho^c(\bm h)&=
\dfrac{2C_{xy}(\bm h)}{C_{x}(\bm 0)+C_{y}(\bm 0)+(\mu-\mu)^2}\\
&= \dfrac{2\rho_{12} \sigma_1 \sigma_2 M(\bm h|\nu_{12},a_{12})}{\sigma_1^2 M(\bm 0|\nu_1,a_1)+\sigma_2^2M(\bm 0|\nu_2,a_2)}\\
&= \dfrac{2\rho_{12} \sigma_1 \sigma_2 M(\bm h|\nu_{12},a_{12})}{\sigma_1^2 +\sigma_2^2}\\
&=\frac{2\sigma_{1}\sigma_2M(\bm h,\nu_{12},a_{12})}{\sigma_1^2+\sigma_2^2}\\
&=\eta \cdot \rho_{12},
\end{align*}
where $\eta=\frac{2\sigma_1\sigma_2 M(\bm h,\nu_{12},a_{12})}{\sigma_1^2+\sigma_2^2}.$

A special case of the Mat\'ern covariance function is when  $\nu_{12}= n + 1/2$. Then 
$$M(\bm h|\nu_{12},a_{12}) = M(\bm h|n+1/2,a_{12}) = \exp(-a_{12} \|\bm h\|) \sum_{k=0}^{n} \dfrac{(n+k)!}{(2n)!} \begin{pmatrix} n\\k \end{pmatrix} (2a_{12} \|\bm h\|)^{n-k},$$ 
and the SCCC is 
$$\rho^c (\bm h)= \dfrac{2\sigma_{12}}{\sigma_1^2 +\sigma_2^2} \exp(-a_{12} \|\bm h\|) \sum_{k=0}^{n} \dfrac{(n+k)!}{(2n)!} \begin{pmatrix} n\\k \end{pmatrix} (2a_{12} \|\bm h\|)^{n-k}. $$
By choosing  $n=0$ and $\nu_{12}=1/2$, $ M(\bm h|1/2,a_{12}) = \exp(-a_{12} \|\bm h\|)$. This gives the SCCC in its simplest form:
$$\rho^c(\bm h)= \dfrac{2\sigma_{12}}{\sigma_1^2 +\sigma_2^2} \exp(-a_{12} \|\bm h\|) .$$
For illustrative purposes, consider  $\sigma_1=1$, $\sigma_2=2$, $\sigma_{12}=1.8$, $a_{12}=1/2$ and $\nu_{12} = \lbrace \frac{1}{2}, \frac{3}{2}, \frac{5}{2} \rbrace$. Then for  $\bm h \in \{0,1, \ldots,15\}$, the patterns of the SCCC for the Mat\'ern covariance function are illustrated in Figure \ref{matern}. The range of the SCCC increases as $\nu$ increases. 
\begin{figure}[h!]
\begin{center}
\includegraphics[scale=0.5]{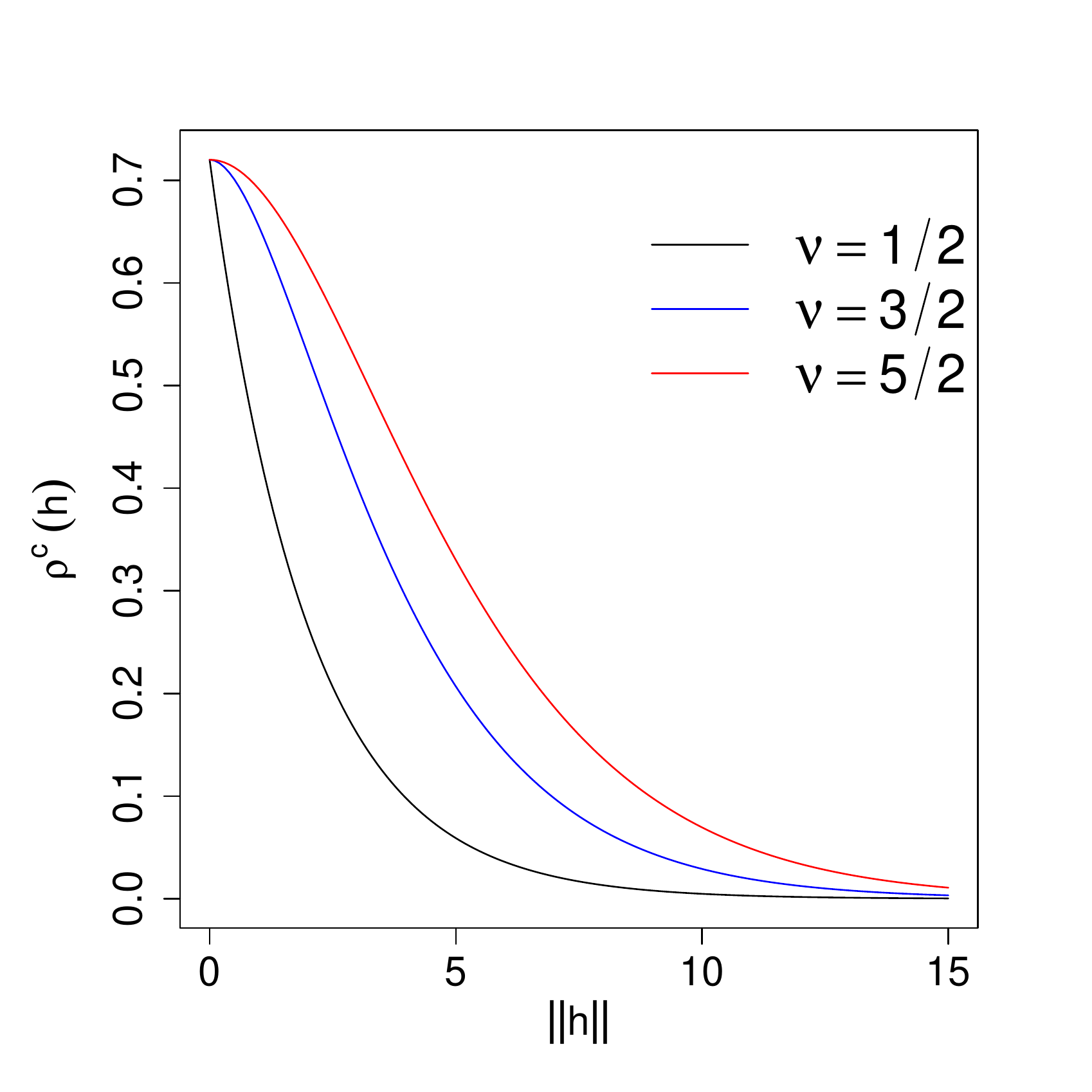}
\end{center}
\vspace{-5mm}
\caption{$\rho^c(\bm h)$ versus $|\bm h|$ for the Mat\'ern covariance function for different values of the smoothness parameter $\nu$.}
\label{matern}
\end{figure}

\item[6.] For a  bivariate Wendland-Gneiting covariance  function \citep{Daley:2015} of the form 
\begin{equation}\label{eq:Wend}
\bm C (\bm h) = \left[ \rho_{ij} \sigma_{ii} \sigma_{jj} R_{ij} (\bm h) \right]_{i,j=1}^2,
\end{equation}
where $R(\bm h, \psi_{12})=c_{ij} b_{ij}^{\nu+2k+1} B(\nu + 2k + 1, \gamma_{ij} +1) \tilde{\psi}_{\nu +\gamma_{ij}+1,k} \left( \dfrac{\| \bm h \|}{b_{ij}} \right), ~ B(\cdot,\cdot)$ is the beta function, and $\tilde{\psi}_{\nu,k} $ is defined, for $k \geq 1$ \citep{Gneiting:2002}, as
\begin{equation*}
\tilde{\psi}_{v,k} (t) = \int_t^1 \dfrac{u(u^2-t^2)^{k-1} (1-u)_+^v}{B(2k,v+1)}du,  \quad 0 \leq t \leq 1
\end{equation*}
the SCCC is 
\begin{equation*}
\rho^c(\bm h)= \dfrac{2 \rho_{12} \sigma_1 \sigma_2  R(\bm h, \psi_{12})}{\sigma_1^2+\sigma_2^2 + (\mu_1 - \mu_2)^2}, \quad \bm h \in \mathbb{R}^2.
\end{equation*}
In particular, considering $R_{ij} (\bm h)=  p_{k}(\| \bm h \|) (1-\| \bm h\| /b_{ij})_{+}^{l}$, where  $k=1$, $l = v + \gamma + 1,$ $\gamma=0$ and $b_{ij}>0$, 
\begin{equation}\label{aux1}
\rho^c (\bm h) = \dfrac{2 \rho_{12} \sigma_1 \sigma_2 \left( 1 + l \| \bm h\|/b_{12} \right) \left( 1 - \| \bm h\|b_{12}\right)_+^l  }{\sigma_1^2 + \sigma_2^2 + (\mu_1-\mu_2)}.
\end{equation}
\end{itemize}
Using similar arguments as in properties 1--6,
the SCCC could be derived for other parametric bivariate correlation functions.

For a bivariate  intrinsically stationary  random field $(X(\bm s), Y(\bm s))^{\top}, ~ \bm s \in \mathbb{R}^2$, with cross-variogram given by
$$\gamma_{XY}(\bm h) = \mathbb{E}[(X(\bm s+\bm h))-X(\bm s)(Y(\bm s+\bm h)-(Y(\bm s))],$$ 
there is another characterizations of the SCCC defined in equation \eqref{eq:spatial_cor}. The coefficient can be
written as
\begin{equation}\label{eq:altern}
\rho^c (\bm h) = 1 - \dfrac{2 \gamma_{xy}(\bm h)}{2 \gamma_{xy}(\bm 0) + (\mu_1-\mu_2)^2 + 2C_{xy}(\bm h)}.
\end{equation}
Because $\rho^c(\cdot)$ in equation \eqref{eq:altern} depends on $C_{XY}(\cdot)$ and $\gamma_{XY}(\cdot)$, we prefer the representation in equation \eqref{eq:spatial_cor}.

\section{Inference}
In the previous section we proved that for several covariance structures, the spatial concordance correlation coefficient defined in equation \eqref{eq:spatial_cor}  can be written as a product of the correlation coefficient and a constant. Thus, we can consider plug-in estimators for the correlation coefficient and the constant. 

Let $(X(\bm s),Y(\bm s))^\top, ~ s\in D\subset \mathbb{R}^2$ be a Gaussian process with mean  $\bm \mu= (\mu_1,\mu_2)^\top$  and covariance function $\bm C(\bm h)$, $\bm s, \bm h \in \mathbb{R}^2$. Then a sample estimate of the SCCC index \eqref{eq:spatial_cor} is 
\begin{equation}\label{eq:concor}
\widehat{\rho}^c (\bm h) = \widehat{\rho}_{12}(\bm h) \widehat{C}_{ab},
\end{equation}
 where $\widehat{C}_{ab} = ((\widehat{a}+1/\widehat{a}+\widehat{b}^2)/2)^{-1}, \quad \widehat{a} = \left( \dfrac{\widehat{C}_{11}(\bm 0)}{\widehat{C}_{22}(\bm 0)} \right)^{1/2}, \quad \widehat{b}= \dfrac{\widehat{\mu_1}-\widehat{\mu}_2}{(\widehat{C}_{11}(\bm 0) \widehat{C}_{22}(\bm 0))^{1/4}},$ and $\widehat{\mu}_1$, $\widehat{\mu}_2,$ $\widehat{C}_{11}(\bm 0)$ and $\widehat{C}_{22}(\bm 0)$ are the maximum likelihood (ML) estimates of $\mu_1$, $\mu_2,$ $C_{11}(\bm 0)$, and $C_{22}(\bm 0)$, respectively.

The asymptotic properties of an estimator as  in  equation \eqref{eq:concor} have  been studied in the literature for specific cases. \cite{Bevilacqua:2015} studied the asymptotic properties of the ML estimator for a  separable Mat\'ern covariance model. They used  a result provided by \cite{Mardia:1984} in an increasing domain sampling framework. Using this theorem and the delta method, we can establish the following result for the Wendland-Gneiting model:
\begin{theorem}\label{main}
Let  $\bm Z(\bm s)=(X(\bm s), Y(\bm s))^\top$, $s\in D\subset \mathbb{R}^2$ be a bivariate Gaussian spatial process with mean $\bm 0$ and covariance function   given by

$$\bm C (\bm h) = \left[ \rho_{ij} \sigma_{ii} \sigma_{jj}  
\left( 1 + (\nu+1) \dfrac{\| \bm h\|}{b_{12}} \right) \left( 1 - \dfrac{\| \bm h\|}{b_{12}}\right)_+^{\nu+1}\right]_{i,j=1}^2,$$
for $\nu>0$ fixed. Define $\bm \theta=(\sigma_1^2,\sigma_2^2, \rho_{12},b_{12})^{\top}$ and denote $\widehat{\bm \theta}_n$ the ML estimator of $\bm \theta.$ Then
$$
\left( \nabla g(\bm \theta)^\top \bm F_n ( \bm \theta)^{-1} \nabla g(\bm \theta) \right)^{-1/2} (g(\widehat{\bm \theta}_n) - g(\bm \theta)) \xrightarrow[]{D} \mathcal{N}(0, 1 ), \ \text{as} \  n\rightarrow \infty,
$$ in an increasing domain sense,
where  $$g(\bm \theta)=  \dfrac{2 \rho_{12} \sigma_1 \sigma_2 \left( 1 + (\nu+1) \dfrac{\| \bm h\|}{b_{12}} \right) \left( 1 - \dfrac{\| \bm h\|}{b_{12}}\right)_+^{\nu+1}  }{\sigma_1^2 + \sigma_2^2},$$  $\bm F_n ( \bm \theta)^{-1}$ is the covariance matrix of  $\widehat{\bm \theta}_n,$
{\footnotesize
 $$\nabla g(\bm \theta) = \begin{pmatrix}
\dfrac{\sigma_2 \rho_{12}(\sigma_2^2-\sigma_1^2) \left( 1 + (\nu +1) \dfrac{\| \bm h\|}{b_{12}} \right) \left( 1 - \dfrac{\| \bm h\|}{b_{12}}\right)_+^{\nu+1}}{\sigma_1 (\sigma_1^2+\sigma_2^2)^2}\\
\dfrac{\sigma_1 \rho_{12}(\sigma_1^2-\sigma_2^2) \left( 1 + (\nu +1) \dfrac{\| \bm h\|}{b_{12}} \right) \left( 1 - \dfrac{\| \bm h\|}{b_{12}}\right)_+^{\nu+1}}{\sigma_2 (\sigma_1^2+\sigma_2^2)^2}\\
\dfrac{2 \sigma_1 \sigma_2 \left( 1 + (\nu +1) \dfrac{\| \bm h\|}{b_{12}} \right) \left( 1 - \dfrac{\| \bm h\|}{b_{12}}\right)_+^{\nu+1}}{\sigma_1^2+\sigma_2^2}\\
\dfrac{2 \sigma_1 \sigma_2 \rho_{12} f(b_{12})}{\sigma_1^2+\sigma_2^2}
\end{pmatrix},$$
}
and
{\small $f(b_{12})=  \left( -\dfrac{(\nu+1) \| \bm h \|}{b_{12}^2} \right) \left( 1- \dfrac{\| \bm h\|}{b_{12}} \right)^{\nu+1}_+ + \left( 1+  \dfrac{ (\nu+1)\|\bm h\|}{b_{12}}\right) \left(1-\dfrac{\|\bm h\|}{b_{12}} \right)^\nu_+ \dfrac{(\nu+1) \| \bm h\|}{b_{12}^2}  .$}%
\end{theorem}
\begin{proof}
See the Appendix.
\end{proof}

\section{A Local Approach}\label{sec:local}

When the size of the images is large, it is difficult to find a single model  fitting  reasonably well to an   entire image.  This has been investigated in the literature for autoregressive processes defined on the plane in the context of image restoration and segmentation. For examples, see \cite{Bustos:2009} and \cite{Ojeda:2010}. 

Here we describe a local approach for a bivariate process of the form  $\bm Z(\bm s) = (\bm Z_1(\bm s),\bm Z_2(\bm s))^\top$, $\bm s \in D \subset \mathbb{R}^2$, where the observations are located over a rectangular grid of size  $n \times m$. The extension to an $l \in \mathbb{N}$-variate process is natural when $l>2$. In this framework, we assume that the whole domain $D$ can be divided into  $p$ sub-windows $D_i$, such that $\cup_{i=1}^n D_i=D$, for $i=1,...p$. Then  we define $p$ processes of the form $\bm Z_i(\bm s) = (\bm Z_{i1}(\bm s), \bm Z_{i2}(\bm s))^\top, ~  \bm s \in D_i$, where each process has a covariance function given by
 
$$\bm C_i(\bm h) = \left[ \rho^i_{jk} \sigma^i_{jj} \sigma^i_{kk} R_i (\bm h, \psi_i) \right]_{j,k=1}^2, \enskip i=1,...p.$$

Then for each local process $\bm Z_i(\cdot)$ we define the local SCCC  $\rho_i^c (\cdot)$ using the theory developed in Section \ref{sec:Spatial_cor}:
\begin{equation}\label{eq:rho_i_c}
\rho^c_i (\bm h) = \dfrac{2 \sigma_{1i} \sigma_{2i}}{\sigma_{1i}^2+\sigma_{2i}^2} \rho_{12i} R_i(\bm h , \psi_i). 
\end{equation}

Based on the local coefficients $\rho_i^c(\cdot)$, we suggest two global  SCCCs. The first one is the average of the $p$ local coefficients, given by
\begin{equation}\label{eq:rho1}\rho_1 (\bm h) = \dfrac{1}{p} \sum_{i=1}^p  \rho_i^c (\bm h). 
\end{equation}
The second one considers the average of each parameter in the correlation function such that the global coefficient is
\begin{equation}\label{eq:rho2}\rho_2 (\bm h) = \dfrac{2 \overline{\sigma}_1 \overline{\sigma}_2}{\overline{\sigma}_1^2+\overline{\sigma}_2^2} \overline{\rho}_{12} R(\bm h ,\overline{\bm \psi}),
\end{equation}
where $\overline{\sigma}_1=\frac{1}{p}\sum_{i=1}^p \sigma_{1i}^2$, and similarly for 
$\overline{\sigma}_2, ~ \overline{\rho}_{12},$ and $\overline{\bm \psi}$. 
As a result we have two global coefficients of spatial concordance depending on averages, the  first one is the average of the local coefficients and the second one is a plug in of the parameter averages. 

When process $\bm Z(\bm s)$ have been observed in the sites $\bm s_1,\ldots,\bm s_n$ and  all the local coefficients have been computed, the sample versions of $\rho_1(\cdot)$ and $\rho_2(\cdot)$ are
\begin{align*}
\widehat{\rho}_1 (\bm h) &= \dfrac{1}{p} \sum_{i=1}^p  \widehat{\rho}^c_i (\bm h)\\ \widehat{\rho}_2 (\bm h) &= \dfrac{2 \widehat{\overline{\sigma}}_1 \widehat{\overline{\sigma}}_2}{\widehat{\overline{\sigma}}_1^2 +\widehat{\overline{\sigma}}_2^2}  \widehat{\overline{\rho}}_{12} R(\bm h ,\widehat{\overline{\bm \psi}}),
\end{align*}
where $\widehat{\overline{\sigma}}_1$, $\widehat{\overline{\sigma}}_2$, $\widehat{\overline{\rho}}_{12}$, and $\widehat{\overline{\bm \psi}}$ 
are the ML estimations of the parameters defined in equation \eqref{eq:rho2}.

Considering and increasing domain sampling scheme, the asymptotic normality of $\widehat{\rho}_1 (\bm h)$ is straightforward. Indeed, let $\bm Z_i(\bm s) = ( Z_{i1}(\bm s), Z_{i2}(\bm s))^\top, ~ s \in D_i$, be a bivariate process with correlation structure given by 
$\bm C_i(\bm h) = \left[ \rho^i_{jk} \sigma^i_{jj} \sigma^i_{kk} R_i (\bm h, \psi_i) \right]_{j,k=1}^2, ~ i=1,...p.$ Define the parameter vector $\bm \theta^i = (\rho_{12}^i,\sigma_1^i,\sigma_2^i,\sigma_{12}^i,\psi_i)^\top$ associated with $\bm Z_i(\bm s)$. If the covariance satisfies the \cite{Mardia:1984} conditions, then
$$\widehat{\bm \theta}^i_n \xrightarrow[]{D} \mathcal{N}(\bm \theta^i, \bm F^i_n(\bm \theta^i)^{-1}),$$ where $\bm F^i_n(\bm \theta^i)$ is the  covariance matrix  of $\widehat{\bm \theta}^i $.
Then for $g(\bm \theta)= \rho_i^c (\bm h)$, we have that
$$ \left(\nabla g (\bm \theta^i)^\top \bm F^i_n(\bm \theta^i)^{-1} \nabla g(\bm \theta^i)\right)^{-1/2} (g(\widehat{\bm \theta^i}_n) - g(\bm \theta^i)) \xrightarrow[]{D} \mathcal{N}(\bm 0, 1).$$

Now assuming that $\bm Z_i(\bm s)$ and  $\bm Z_j(\bm s)$ are independent for all $i \neq j$, we get
$$ \widehat{\rho}_1(\bm h) = \dfrac{1}{p} \sum_{i=1}^p \widehat{\rho_i}^c (\bm h)  \xrightarrow[]{D} \mathcal{N}\left( \rho_i^c (\bm h), \dfrac{1}{p^2} \sum_{i=1}^p \nabla g (\bm \theta^i)^\top \bm F^i_n(\bm \theta^i)^{-1} \nabla g(\bm \theta^i) \right).$$

\section{Monte Carlo Simulations}\label{sec:sim}
We used Monte Carlo simulation to explore the properties of the SCCC, $\rho^c(\cdot)$, for finite samples sizes. The performance of the ML estimations were then analyzed with respect to the true values of the coefficient. We generated 500 replicates from a Gaussian random field sampled on a regular lattice of size $20\times 20$ inside the region $[-\frac{3}{2}, \frac{3}{2}]^2.$ Each replicate was generated from a bivariate Gaussian random field with mean zero and Wendland-Gneiting covariance function given in equation \eqref{eq:Wend}. In each case, we estimated the parameters of the covariance function using ML and used them to compute the SCCC given in equation \eqref{eq:Wend}. Three set of parameters were considered, one set each for $\sigma_1=\sigma_2=1$, $\nu=4$ and $k=1$:

\begin{enumerate}
    \item Case 1: $\rho_{12} = -0.15$, $b_{1}=0.5$, $b_2=0.4$, and $b_{12}=0.35$.
    \item Case 2: $\rho_{12} = 0.25$, $b_{1}=1.2$, $b_2=0.9$, and $b_{12}=1$.
    \item Case 3: $\rho_{12} = 0.3$, $b_{1}=1.8$, $b_2=1.4$, and $b_{12}=1.5$.
\end{enumerate}

\begin{figure}[H]
\begin{minipage}[c]{0,5 \textwidth }
\begin{center} 
\includegraphics[scale=0.35,trim=0 0 0 0,clip]{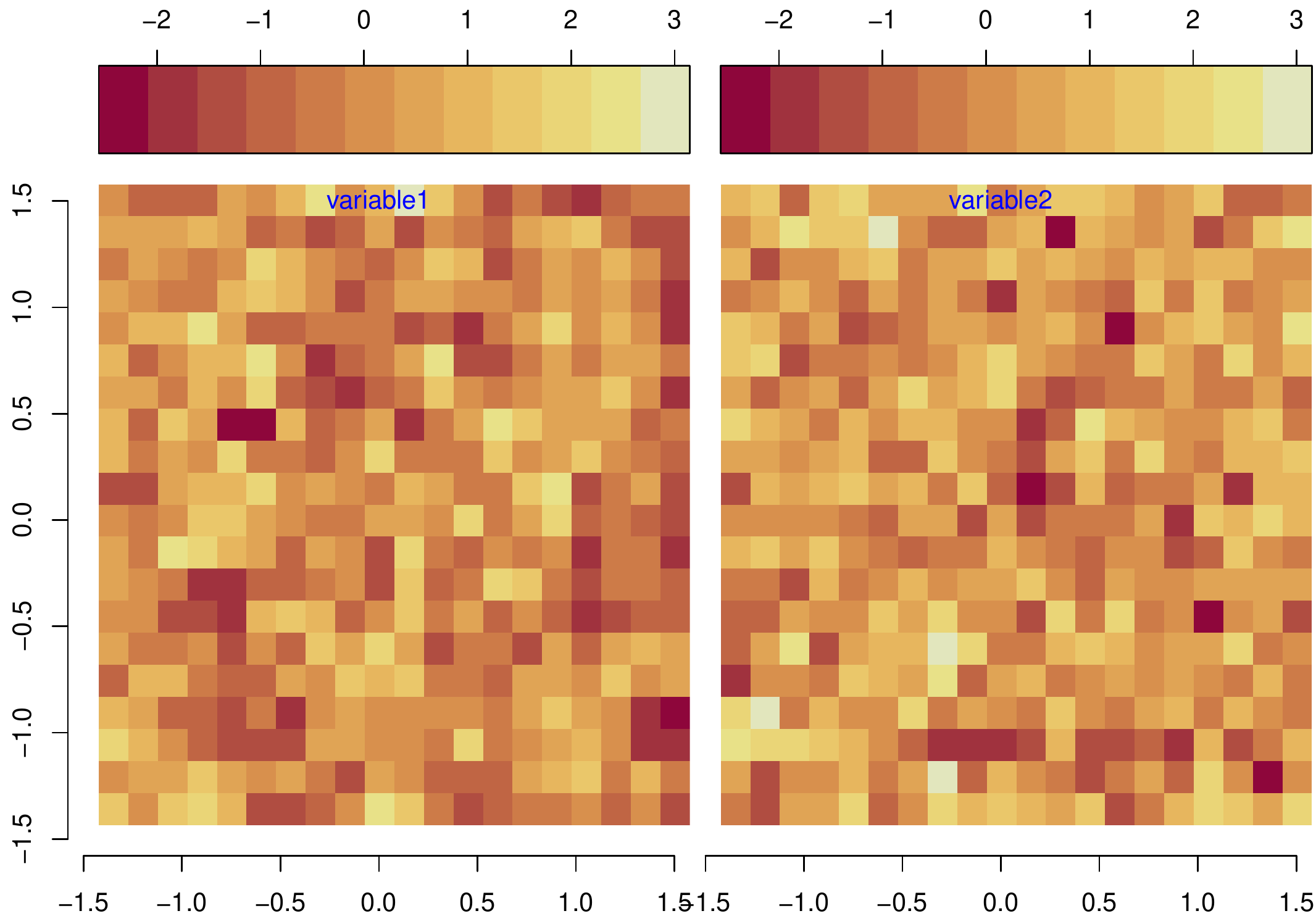} \\
(a)\end{center}
\end{minipage}
\begin{minipage}[c]{0,5 \textwidth }
\begin{center}
\includegraphics[scale=0.35,trim=0 0 0 0,clip]{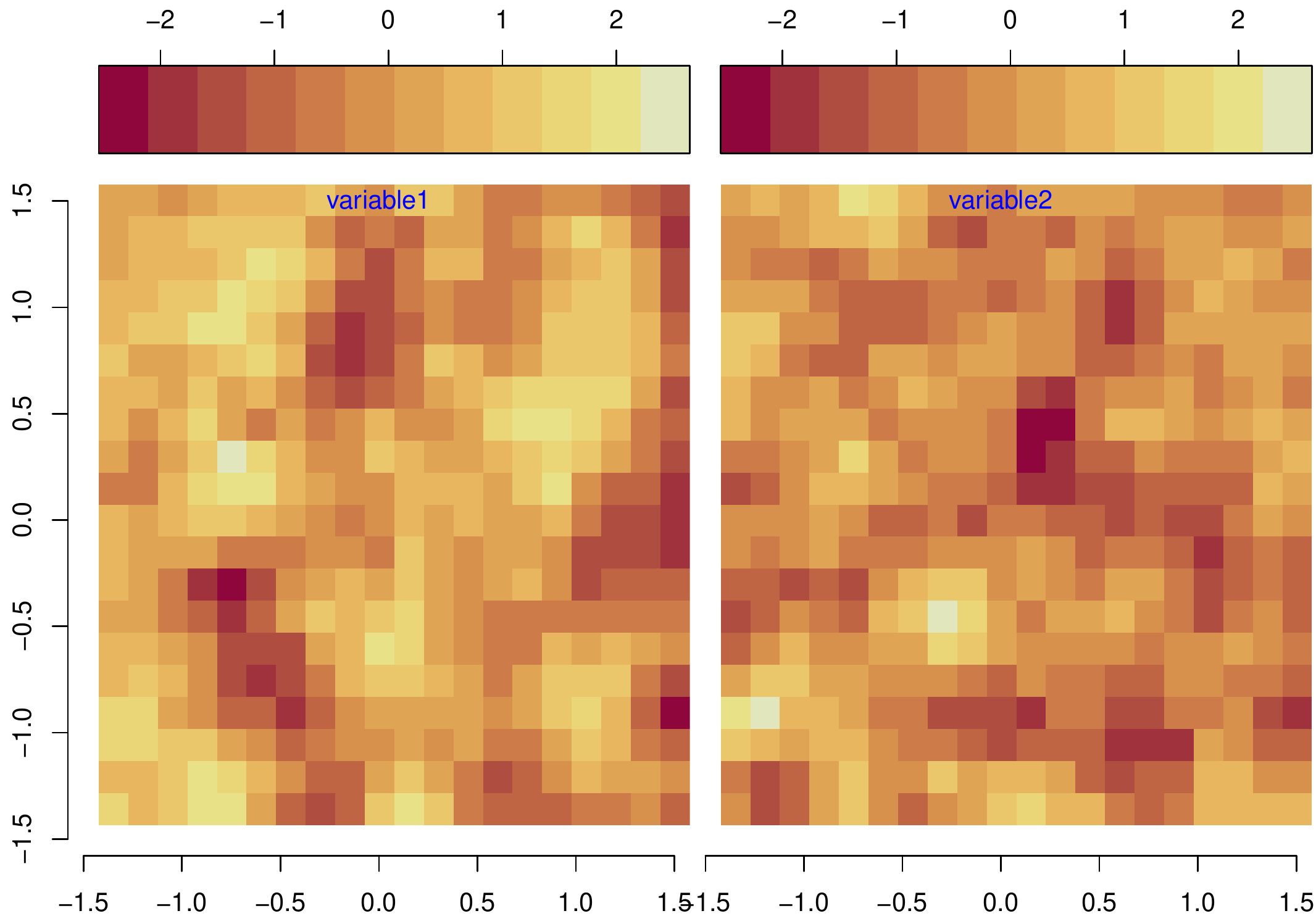} \\
 (b)\end{center}
\end{minipage}
\begin{center}
\begin{minipage}[c]{0,5 \textwidth }
\begin{center}
\includegraphics[scale=0.35,trim=0 0 0 0,clip]{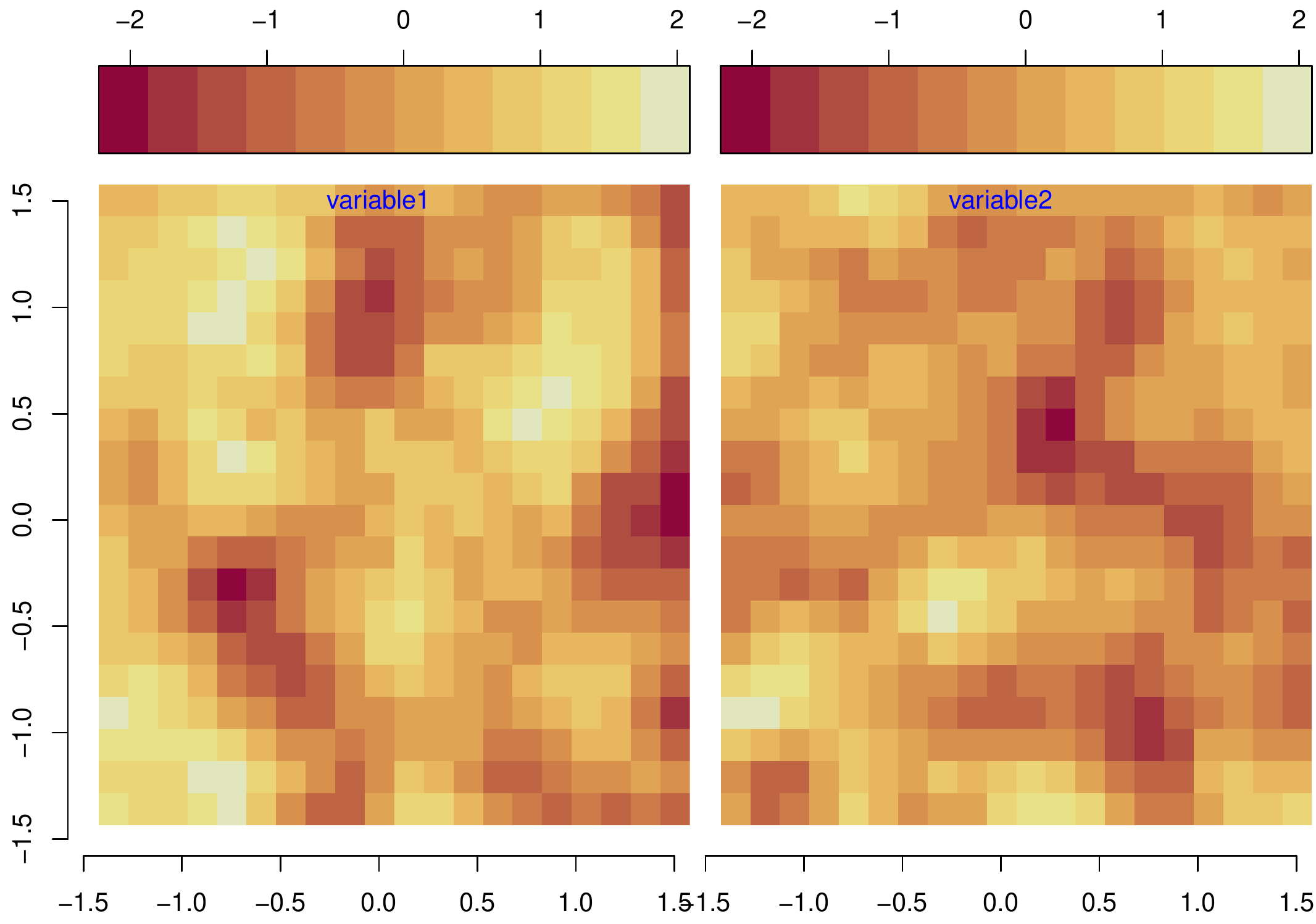} \\
 (c)\end{center}
\end{minipage}
\end{center}
\caption{Realization of a Gaussian random field with  bivariate Wendland-Gneiting correlation function. (a) Case 1; (b) Case 2;  (c) Case 3.}
\label{simuwendland1}
\end{figure}

In Figure \ref{simuwendland1} we show a realization of the random field for each case.

\begin{figure}[H]
\begin{center} 
\subfigure[]{\includegraphics[scale=0.35]{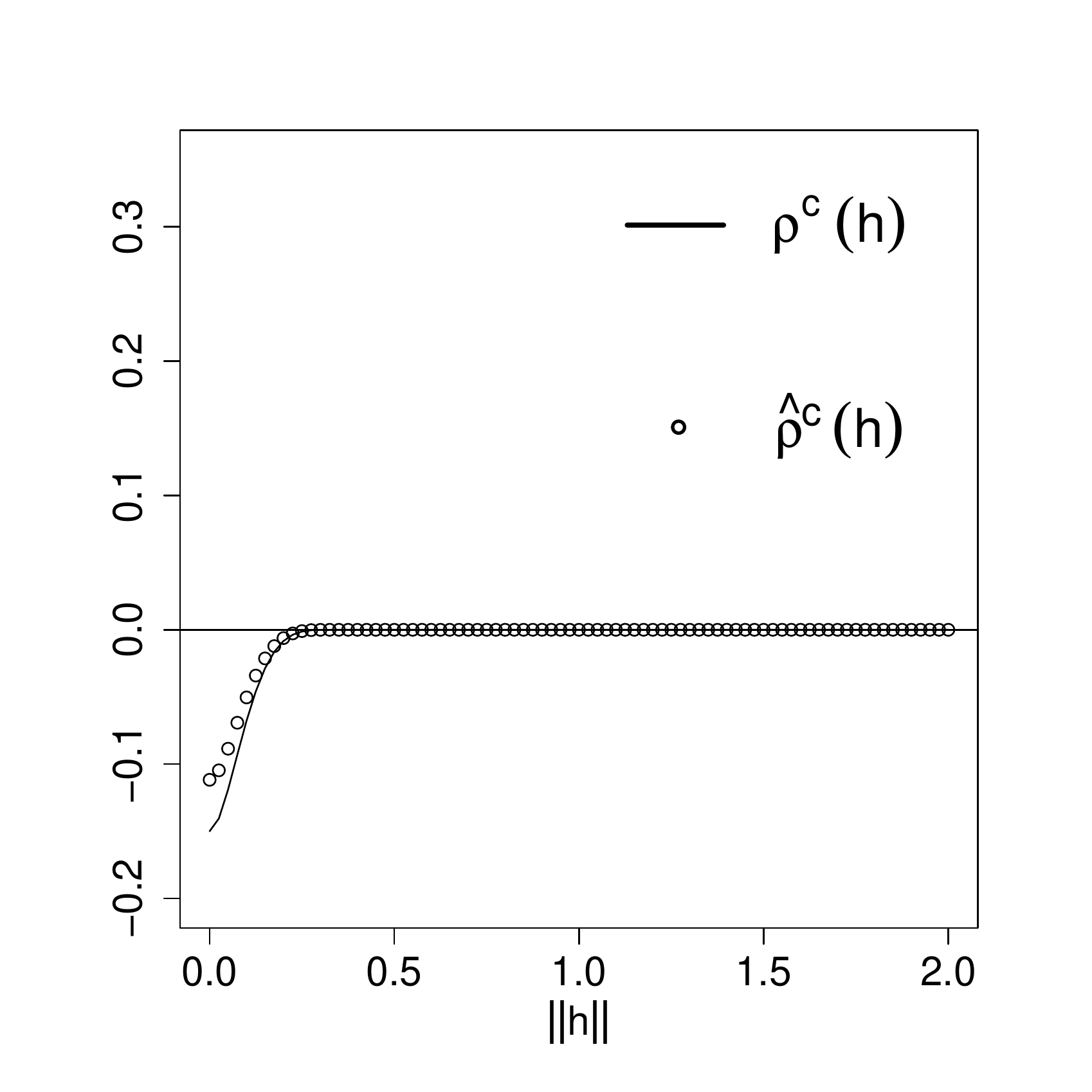} }
\subfigure[]{\includegraphics[scale=0.35]{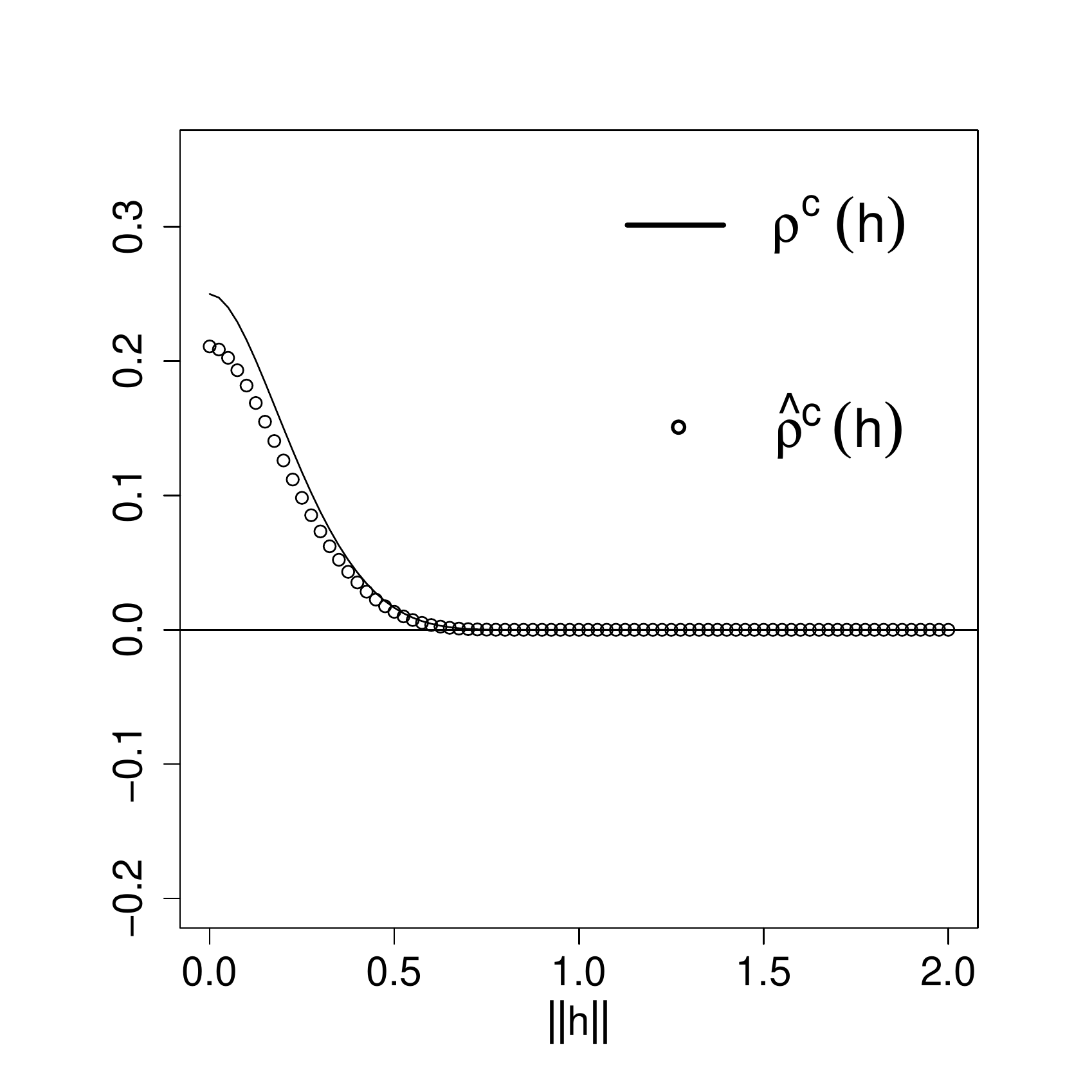} }\\
\subfigure[]{\includegraphics[scale=0.35]{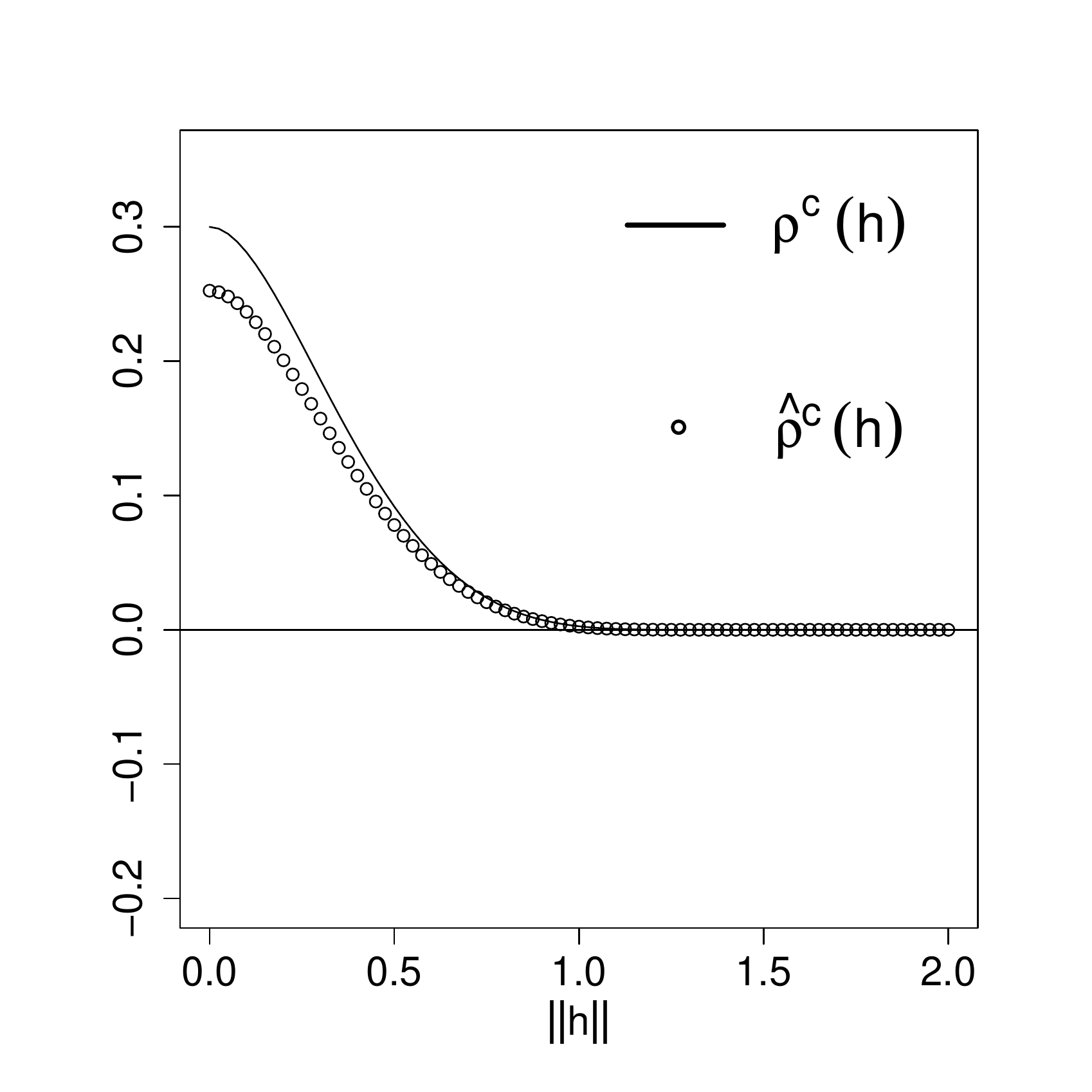}} 
\end{center}
\caption[Coeficiente de concordancia teórico y estimado para covarianza de Wendland-Gneiting.]{Theoretical coefficients (solid lines) and estimates (circles) for the distinct sets of parameters. (a) Case 1; (b) Case 2; (c) Case 3.}
\label{compsimuwend}
\end{figure}

The ML estimates of the parameters of the Wendland-Gneiting covariance function had low bias and standard errors, and agreed with previously published results (e.g., \citeauthor{Bevilacqua:2019}, \citeyear{Bevilacqua:2019}). Using these estimates, we computed the SCCC in each case for $0<||\bm h||<2$. The mean square errors of the estimates were bounded by $3.9404 \cdot 10^{-5}$, $9.7958 \cdot 10^{-5}$, and $0.0002$, respectively, for cases 1--3.  $\rho^c(\bm h)$ \textit{versus} $||h||$ and $\widehat{\rho}^c(\bm h)$ \textit{versus} $||h||$ are plotted in Figure \eqref{compsimuwend}; the true coefficient is drawn with a continuous line. The estimates of the SCCC were reasonably well-behaved but worsened when $||h||$ was close to zero, as is typical of lag-dependent spatial functions computed over a rectangular grid where the minimum distance between coordinates is fixed. 
The general Monte Carlo simulation study also involved the bivariate Mat\'ern covariance function and the results were similar. The estimate of $\rho_{12}$ was better for the Mat\'ern case in terms of the mean square error. This is important because in both cases, the estimate of $\rho_{12}$ affected the estimate of the SCCC. 

Finally, for the same region used in the previous Monte Carlo simulation, we computed the asymptotic variance of $\widehat{\rho}^c(\cdot)$. For $0<||\bm h||<2$, all variances are bounded by 0.006, and the largest discrepancies between cases 1--3 were seen near the origin.

To gain more insight into the computational time required for computing $\widehat{\rho}^c(\cdot)$ for the covariance functions used in this work, we ran similar simulations with different window sizes. We ran 100 simulations, and in each, $\widehat{\rho}^c(\cdot)$ was computed for the Mat\'ern and Wendland-Gneiting covariance functions for window sizes = $8\times 8,$ $12\times 12$, $16\times 16$, and $20\times 20$. All computations were done using an HP ProLiant DL380G9 server, equipped with a 2x Intel Xeon E5-2630 v3 2.40 GHz processor, 128 GB DDR4 2.133 Ghz RAM, and 512 GB SSD storage.

\begin{figure}[H]
\begin{center}
\includegraphics[scale=0.5]{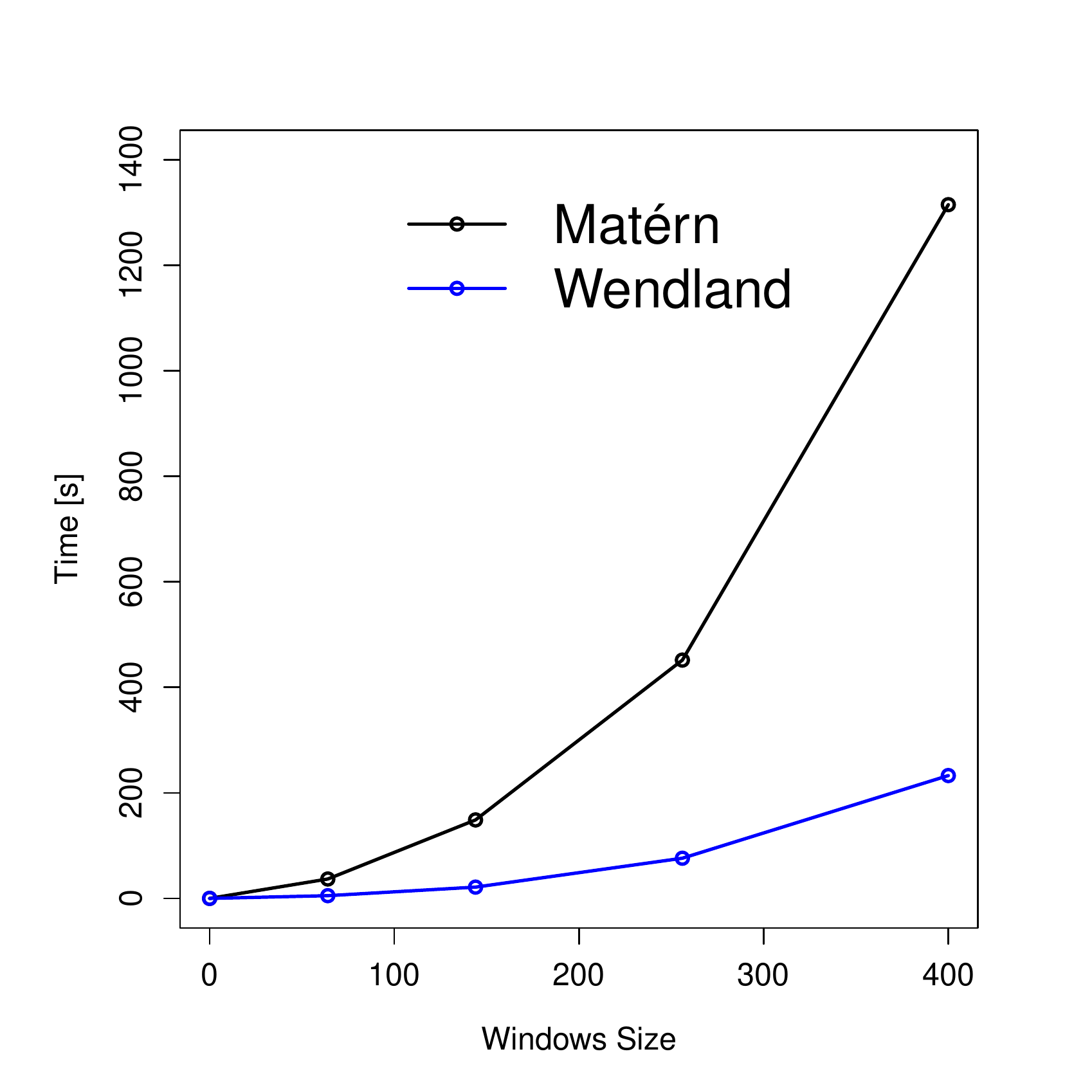}
\end{center}
\vspace{-25pt}
\caption{Computational time in seconds to compute $\widehat{\rho}^c(\cdot)$ for the Mat\'ern and Wendland-Gneiting covariance functions.}
\label{fig:time}
\end{figure}

Time to run each simulation increased exponentially with window size (Figure \ref{fig:time}). Although the time required to compute the Wendland-Gneiting covariance function was always smaller than the time to compute the Mat\'ern covariance function, for real images it is not feasible to compute $\widehat{\rho}^c(\cdot)$, at least using an interpreted language like \R \ as we did here. This result further supports the use of the local approach we presented in Section \ref{sec:local}, but we will continue to explore ways to optimize and accelerate the computation of $\widehat{\rho}^c(\cdot)$.

\section{An Application}
\subsection{Motivation}
Our application derives from ecology.  In order to track the seasonality (``phenology'') of vegetation activity in different ecosystems, digital cameras have been deployed to record high-frequency images of the canopy at hundreds of research sites  around the world \citep{Richardson:2019}. From each image, color-channel information  (e.g., RGB [red-green-blue] values of each pixel) are extracted and converted to a suite of ``vegetation indices'' derived from linear or nonlinear transformations of the RGB or other color spaces \citep{Sonnentag:2012,Mizunuma2014,Toomey2015, NguyRobertson2016}. These indices have been used to identify the timing of seasonal phenomena such as leaf-out, senescence, and abscission, and to monitor how these phenomena are changing in response to ongoing climatic change \citep{Sonnentag:2012}. However, different cameras may render the same scene differently because of the specifics of the imaging sensor being used (e.g., CCD, CMOS) and researchers have used a wide range of different cameras because of considerations including trade-offs between cost and image quality. Additionally, changes in scene illumination (e.g., caused by time-of-day or cloud cover) also may impact the resulting image. Although previous research has shown that diurnal, seasonal, and weather-related changes in illumination can have large effects on estimates of average color (or color index) for the whole image or a region of interest \citep{Sonnentag:2012}, spatial information has not been incorporated previously in these estimates.  

\subsection{Imagery}

\begin{figure}[H]
\begin{center}
 \subfigure[]{\label{cam1-orig}\includegraphics[width=7cm]{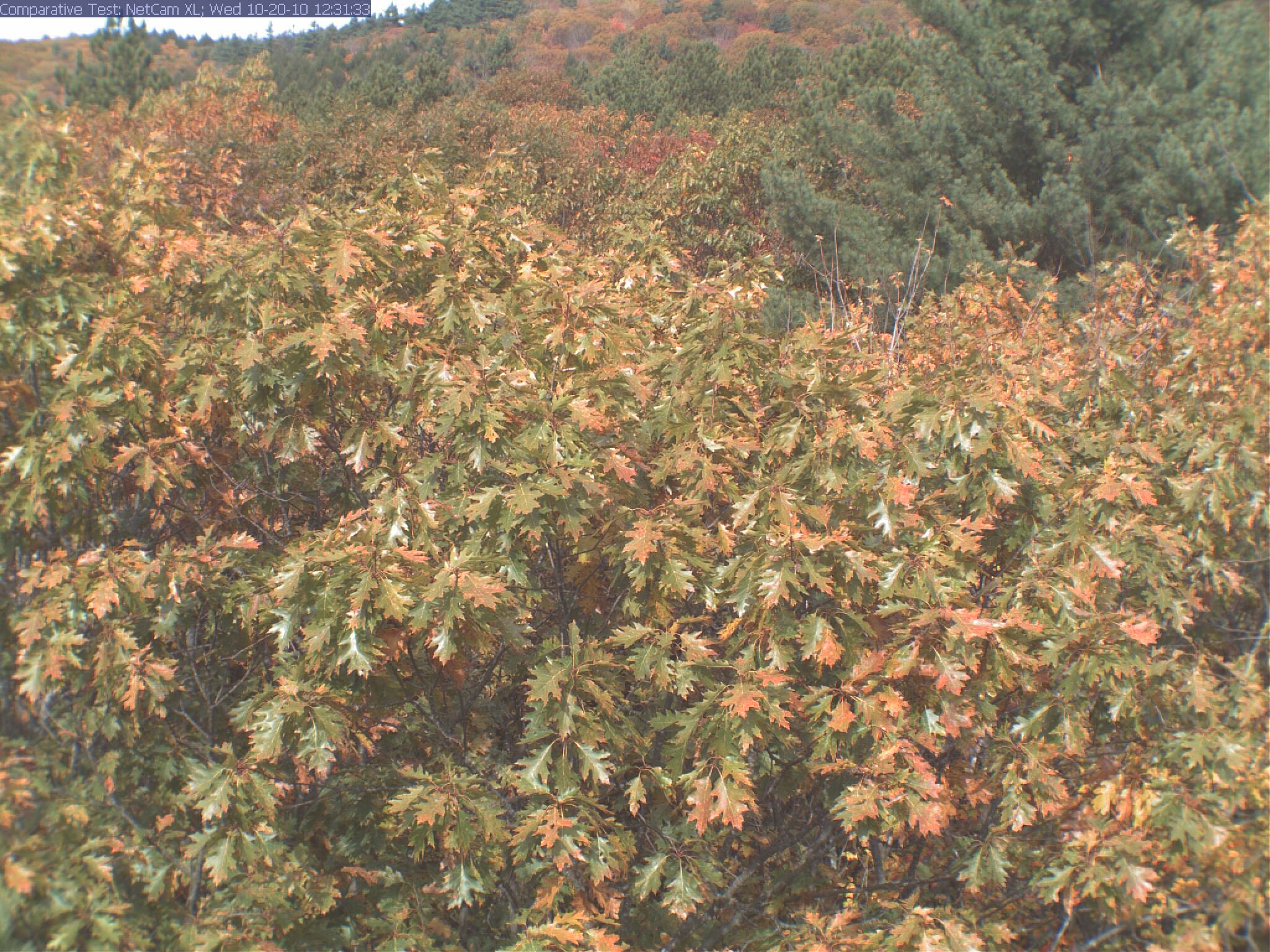}}
\subfigure[]{\label{cam2-orig}\includegraphics[width=7cm]{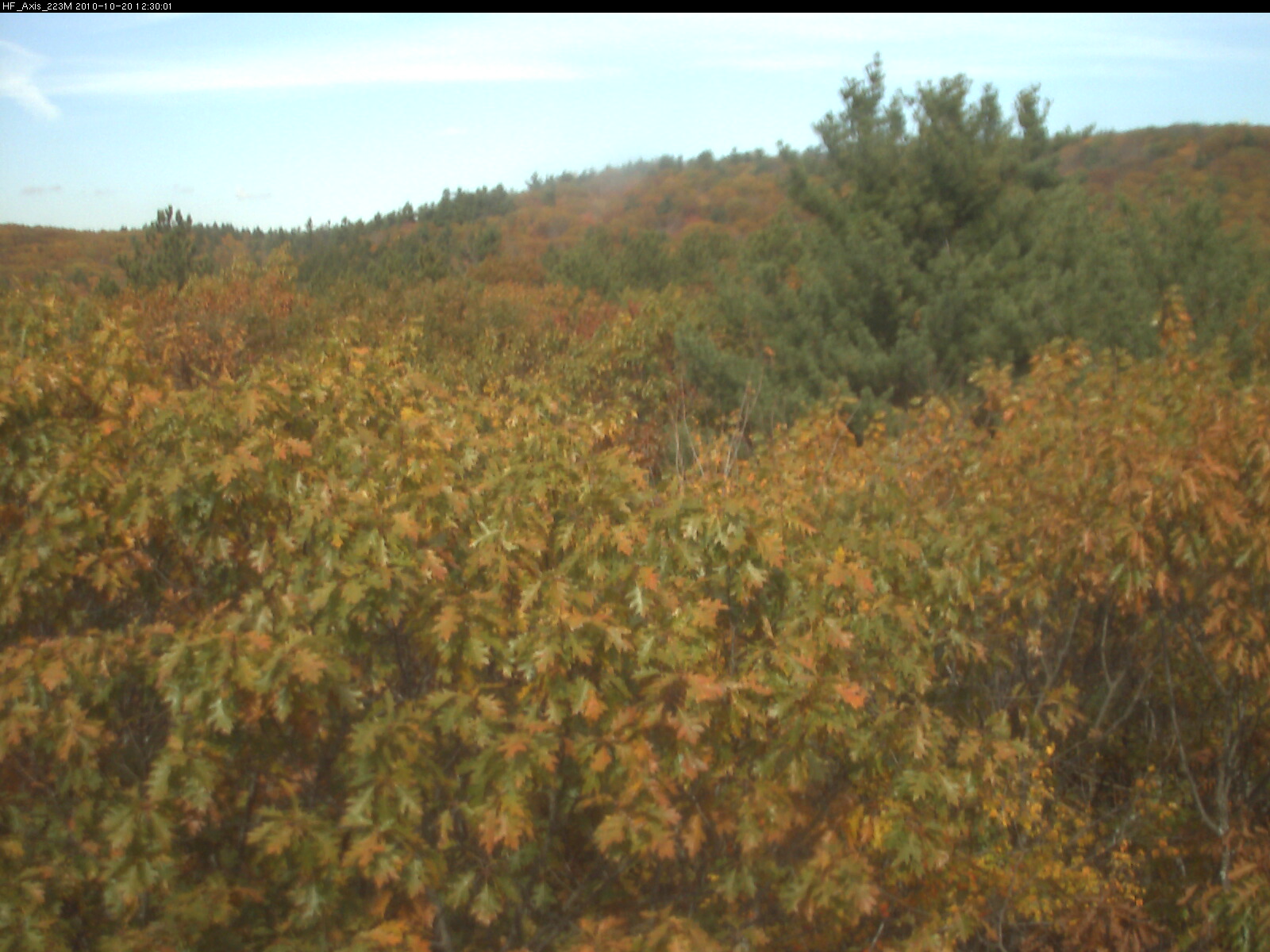}}
\end{center}
\caption{ \scriptsize Two images taken by adjacent cameras of the same site at Harvard Forest. (a): Image taken with an outdoor StarDot NetCam XL 3MP camera. (b): Image taken with an outdoor Axis 223M camera. The dominant tree species (foreground) is red oak (\textit{Quercus rubra}), and there is some white pine (\textit{Pinus strobus}) in the upper right corner.}
\end{figure}

We focus here on comparing two jpeg images taken of the same scene on 20 October 2010 by two different cameras (Figures \ref{cam1-orig}, \ref{cam2-orig}). These images were taken with, respectively, an outdoor StarDot NetCam XL 3MP camera with a $2048 \times 1636$-pixel CMOS sensor (Figure \ref{cam1-orig}) and an outdoor Axis 223M camera with a $1600 \times 1200$-pixel CCD sensor (Figure \ref{cam2-orig}). These images were selected from the image archive associated with an experiment, analyzed and reported on previously by \cite{Sonnentag:2012}, in which images, color time series, and phenological transition dates from eleven different cameras were compared. Although the two images we use here are of the same scene and were taken at the same time, they are not identical. For example, both cameras were pointing due north with an $\approx 20\degree$ tilt angle, but image displacement occurred because the cameras were mounted at different positions on a fixed platform. The resolution and overall field-of-view also differed because of different sensor sizes and lens characteristics. \cite{Sonnentag:2012} compared color information averaged across a small ``region of interest'' in the images. Here, we work with the entire images after correction for differences of field-of-view and displacement.

\begin{figure}[H]
\begin{center}
 \subfigure[]{\label{cam1}\includegraphics[width=7cm]{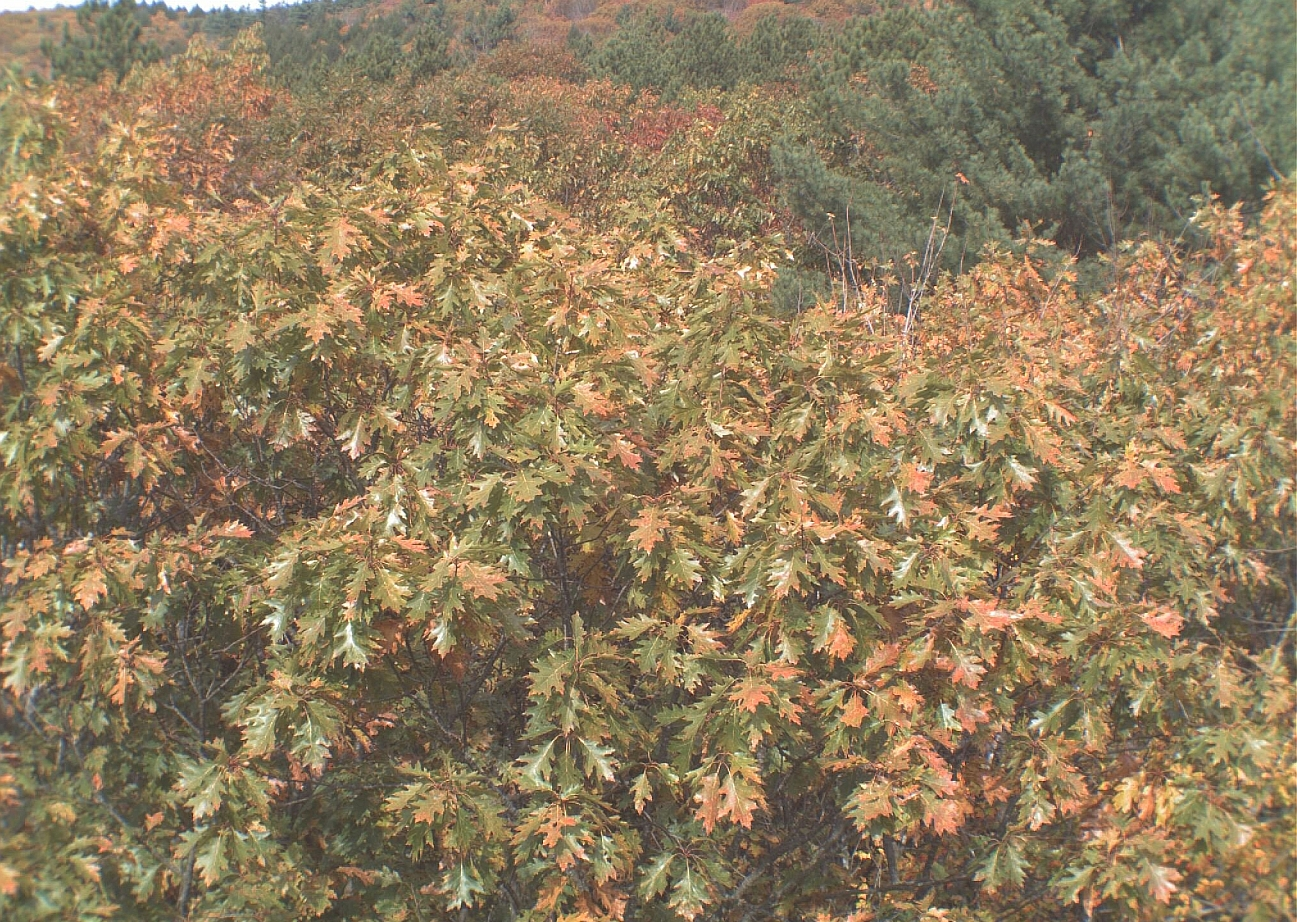}}
\subfigure[]{\label{cam2}\includegraphics[width=7cm]{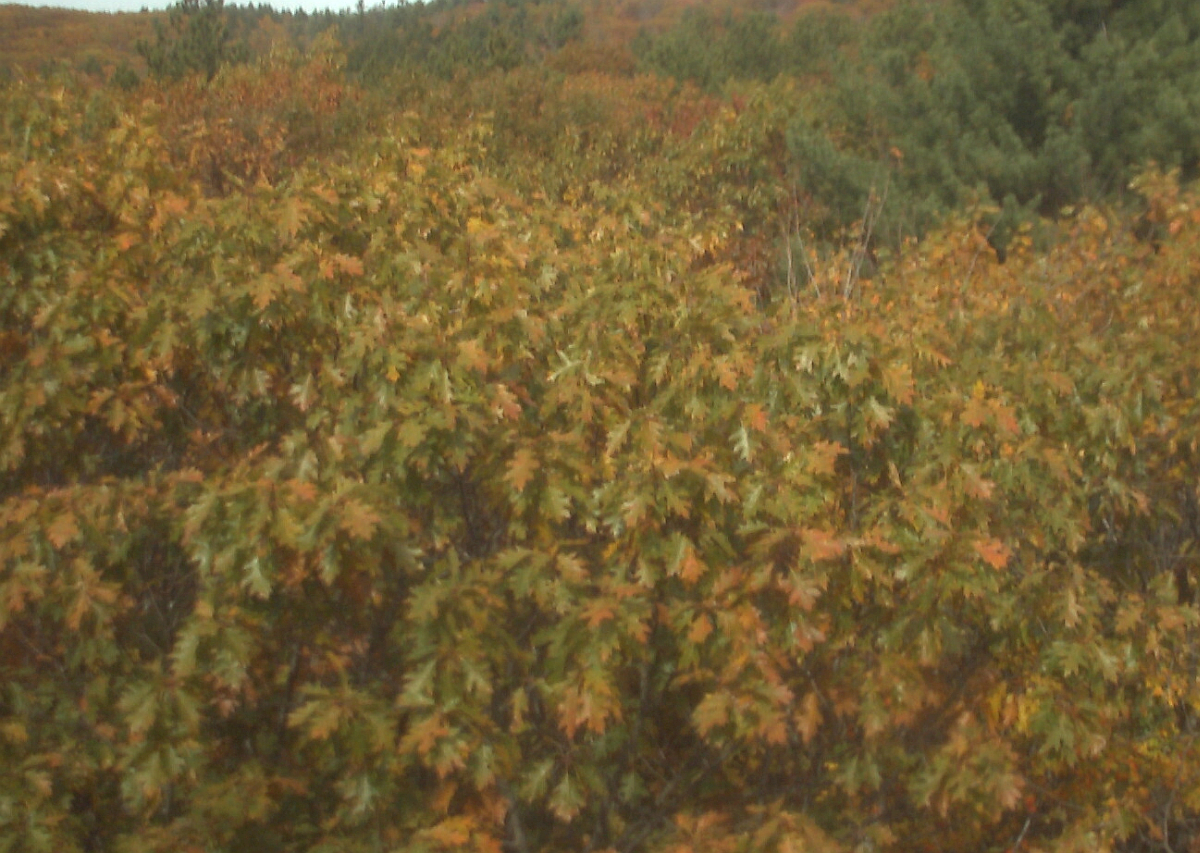}}
\end{center}
\caption{ \scriptsize The two images from Harvard Forest after cropping to equivalent views and resampling to equivalent pixel dimensions. (a): Image taken with an outdoor StarDot NetCam XL 3MP camera (Figure \ref{cam1-orig}; (b): Image taken with an outdoor Axis 223M camera (\ref{cam2-orig}.}
\end{figure}

\begin{figure}[H]
\begin{center}
 \subfigure[]{\label{im1}
 \includegraphics[width=5cm]{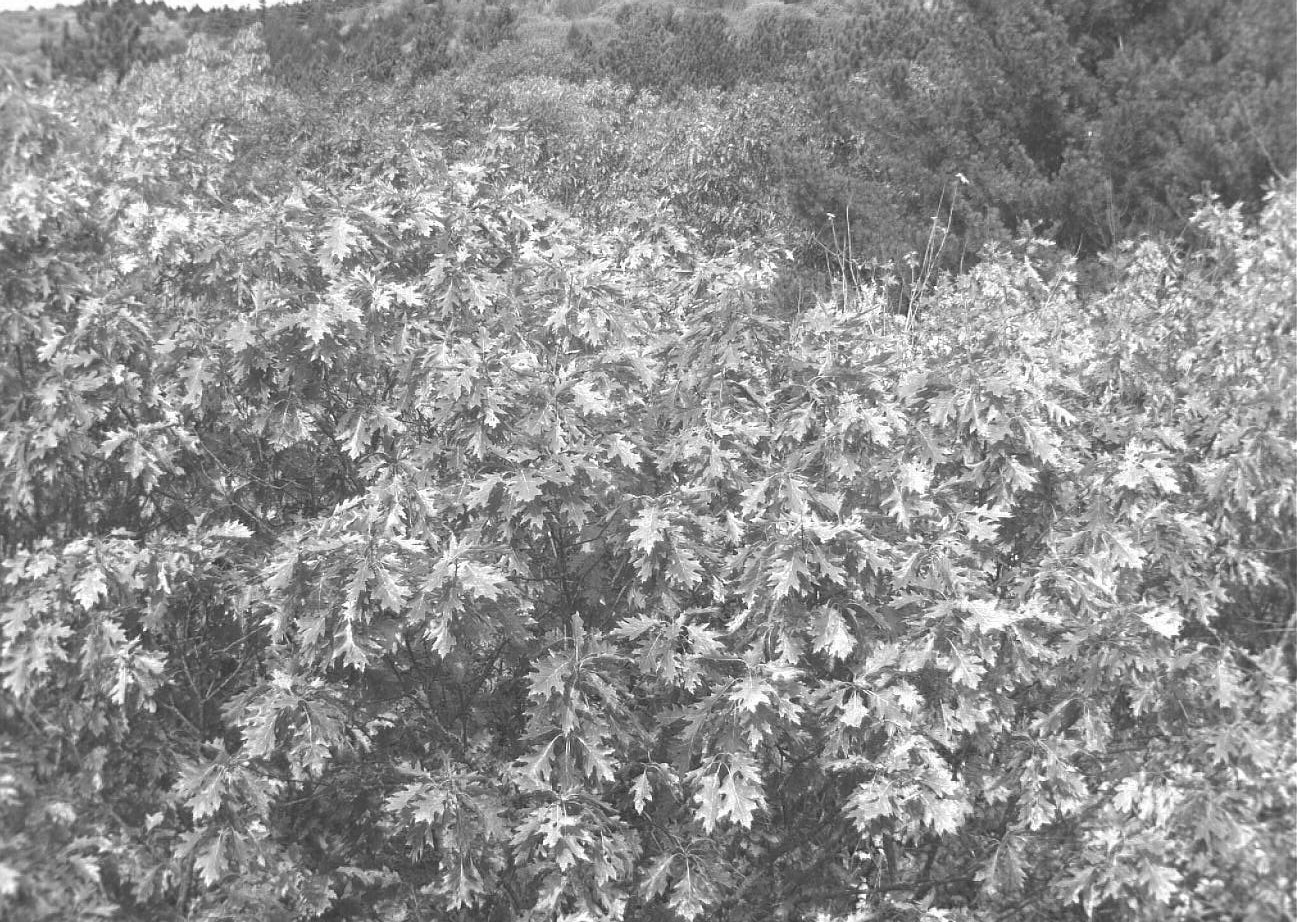}}
 \subfigure[]{\label{im2}\includegraphics[width=5cm]{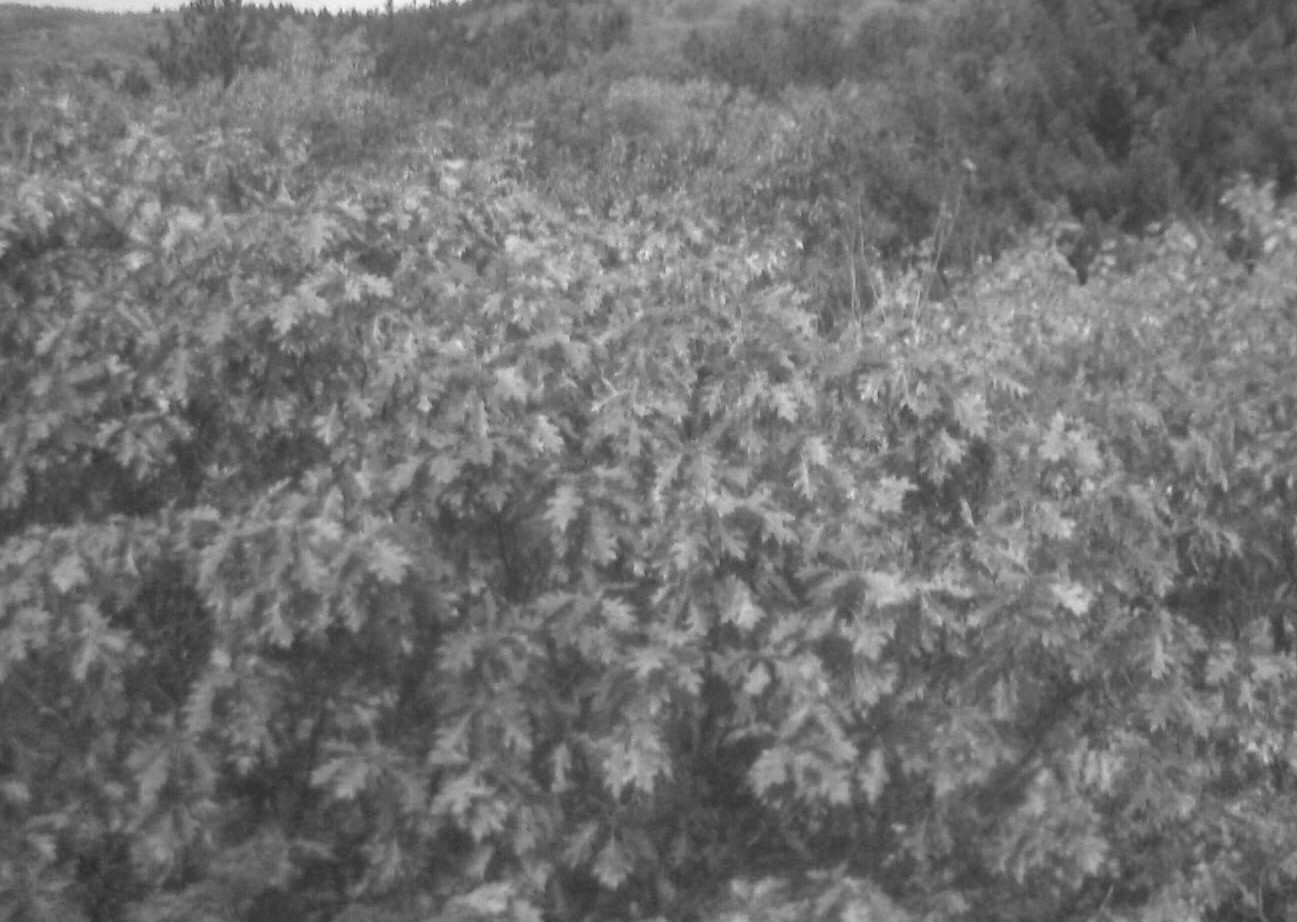}}\\
   \subfigure[]{\label{im2t}\includegraphics[width=5cm]{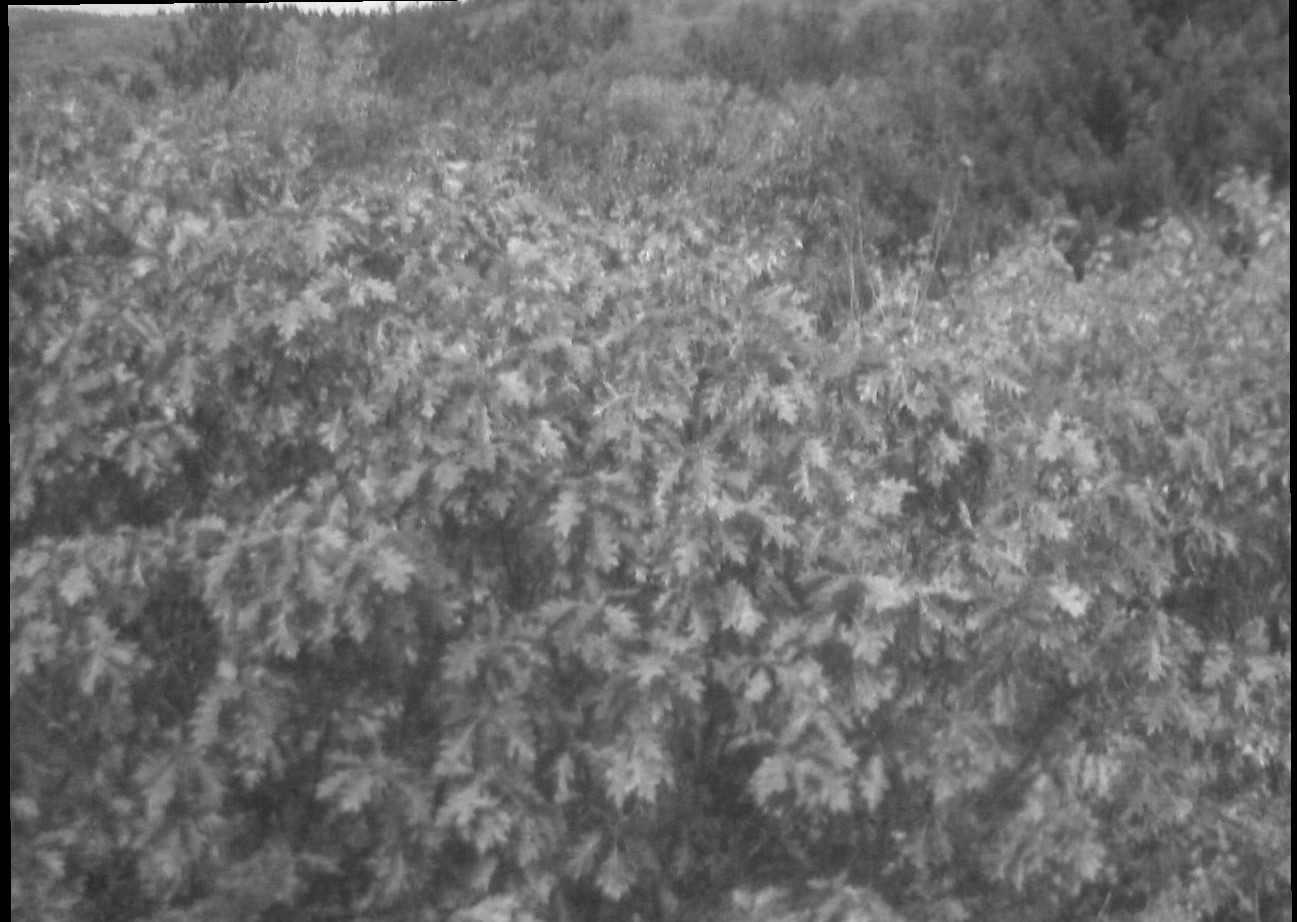}}
\end{center}
\caption{ \scriptsize The two cropped and resampled images (Figures \ref{cam1}, \ref{cam2}) converted to grayscale, and the coordinate registration of the second image with respect to the first. (a): Image taken with an outdoor StarDot NetCam XL 3MP camera (Figure \ref{cam1}); (b): Image taken with an outdoor Axis 223M camera (Figure \ref{cam2}); (c): Image (b) registered to image (a). }
\end{figure}

\begin{figure}[H]
\begin{center}
 \subfigure[]{\label{im1gcc}
 \includegraphics[width=5cm]{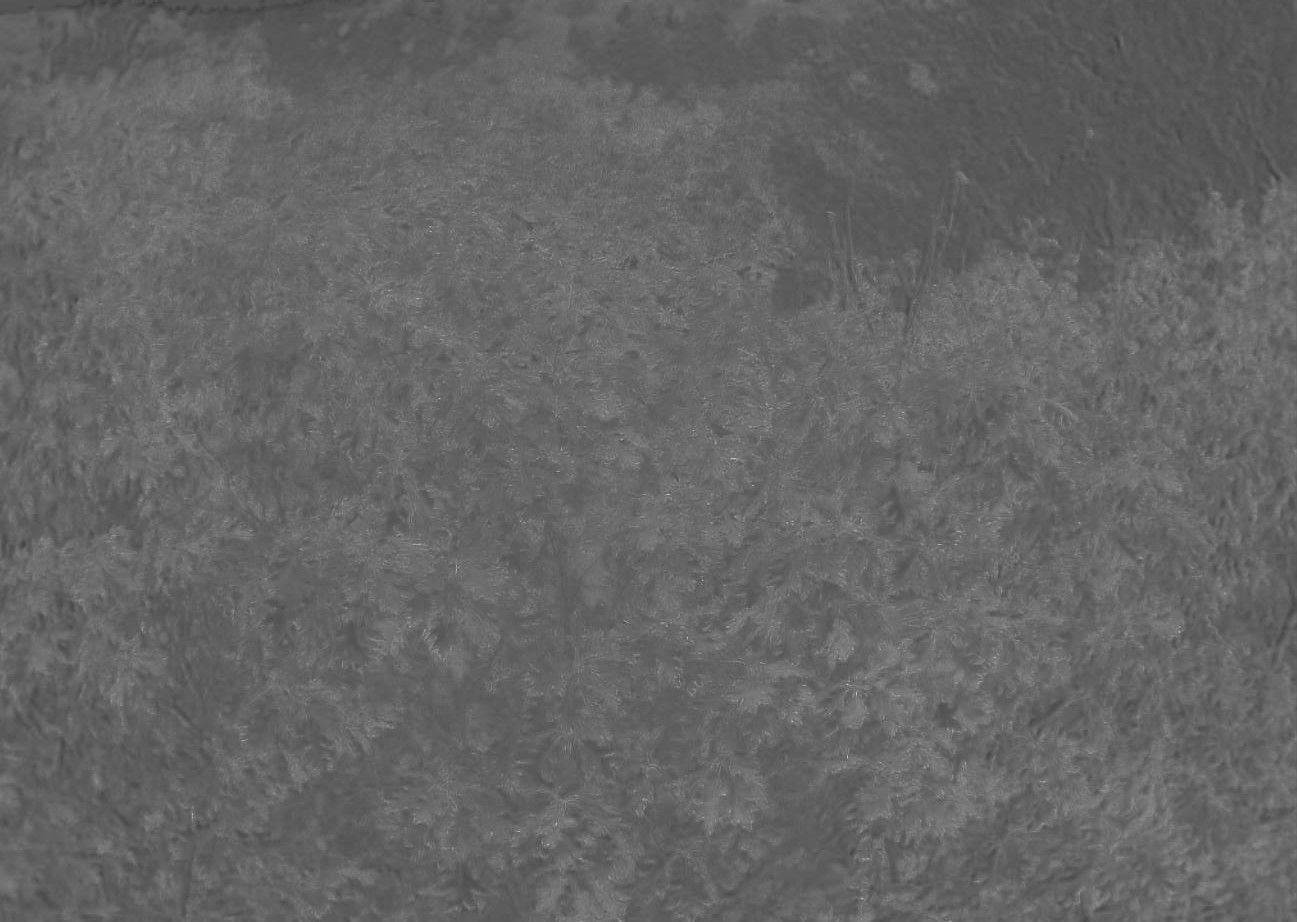}}
 \subfigure[]{\label{im2gcc}\includegraphics[width=5cm]{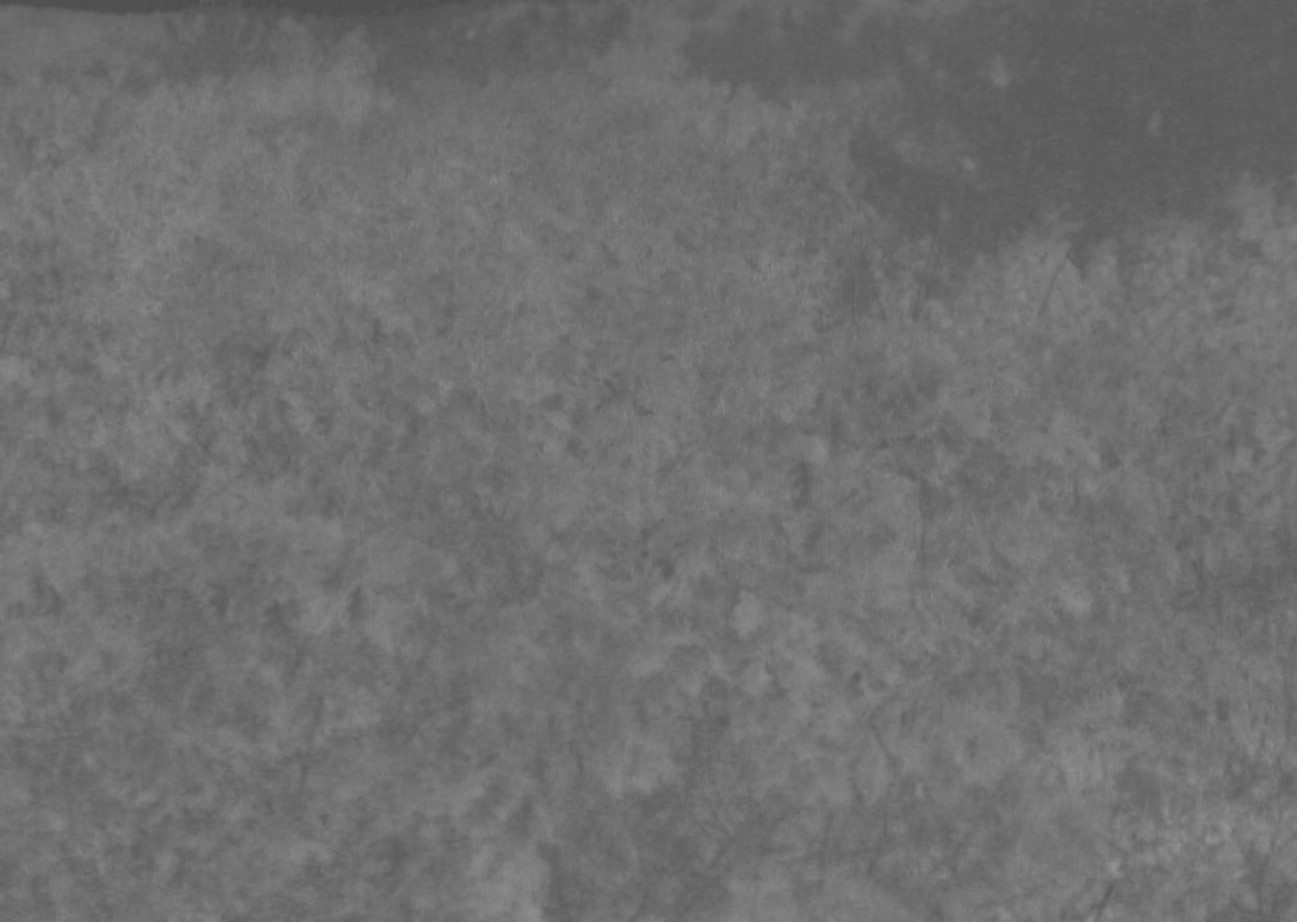}}\\
  \subfigure[]{\label{im2gcct}\includegraphics[width=5cm]{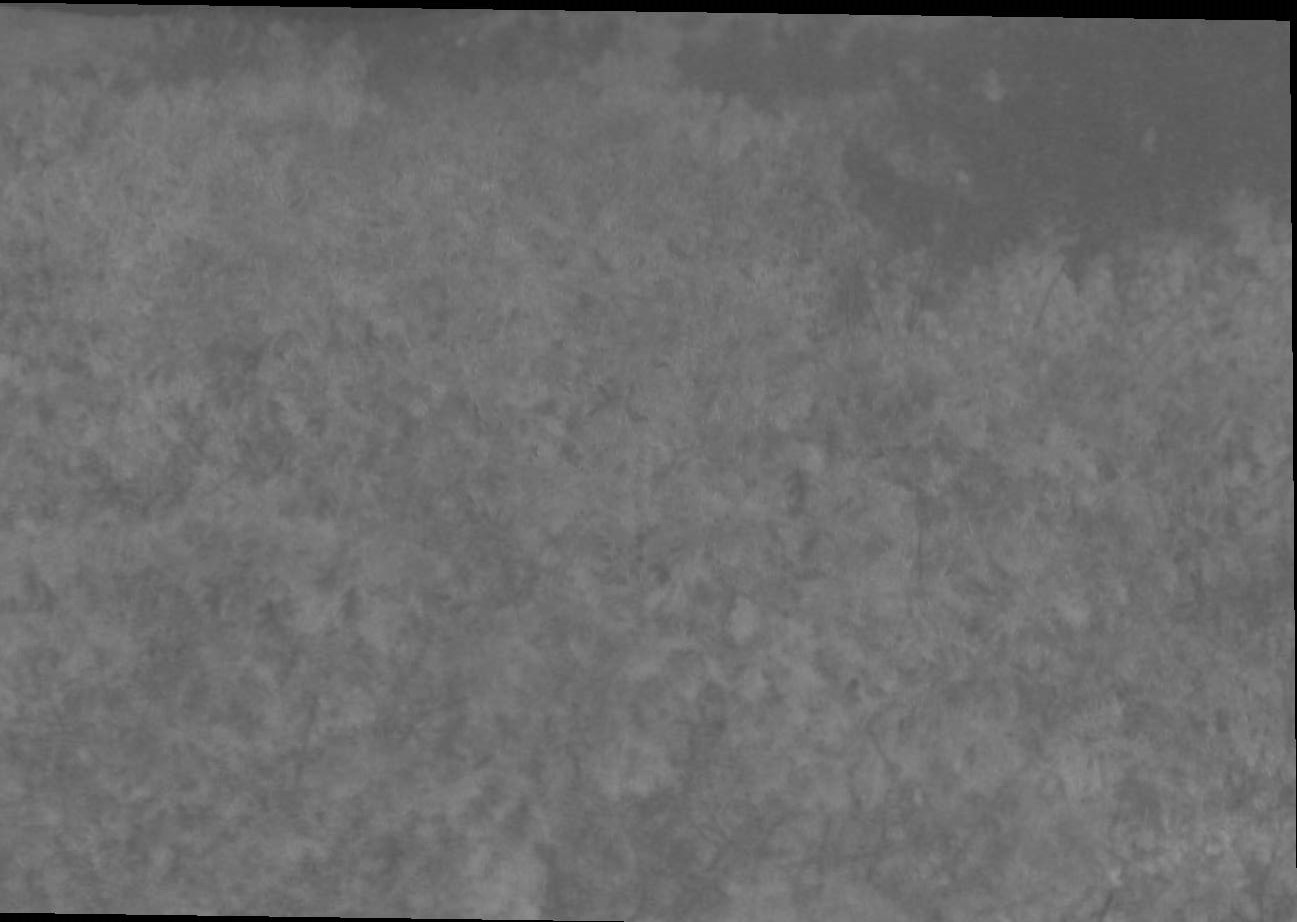}}
\end{center}
\caption{ \scriptsize Cropped and resampled images (Figures \ref{cam1}, \ref{cam2}) corrected for brightness using the green chromatic coordinate (\textit{g\textsubscript{cc}}), and the coordinate registration of the second image with respect to the first. (a): \textit{g\textsubscript{cc}} for Figure \ref{cam1}; (b): \textit{g\textsubscript{cc}} for Figure \ref{cam2}; (c): Image (b) registered to image (a).}
\end{figure}

To account for differences in field-of-view and displacement, the two images were first manually cropped using tools in IrfanView (version 4.38; \citeauthor{Skiljan:2014} \citeyear{Skiljan:2014}) to equivalent areas and aspect ratios. The resulting images had $2023 \times 1444$ pixels for the higher-resolution one taken with the StarDot camera and $1297 \times 922$ pixels for the lower-resolution one taken with the Axis camera. The higher-resolution image was then resized and down-sampled in IrfanView so that it had the same number of pixels as the lower-resolution image (Figures \ref{cam1}, \ref{cam2}). These two images were loaded into the \R \ software package (version 3.51; \citeauthor{R:2018}, \citeyear{R:2018}) using the \texttt{load.image} function in the \texttt{imager} package \citep{Urbanek:2014} and transformed either to gray-scale using the \texttt{grayscale} function in the same package (Figures \ref{im1}, \ref{im2}) or to green chromatic coordinates (\textit{g\textsubscript{cc}}), which normalizes for brightness ($g_{cc}=\frac{G}{R+G+B}$; \citeauthor{Gillespie:1987}, \citeyear{Gillespie:1987}) (Figures \ref{im1gcc}, \ref{im2gcc}). For both the gray-scale and \textit{g\textsubscript{cc}} images, the lower-resolution image (Figures \ref{im2} and \ref{im2gcc}, respectively) was then coordinate-registered to the higher-resolution image (Figure \ref{im1} and \ref{im2gcc}, respectively) using the \R \ package \texttt{RNiftyReg} and a linear (affine) transformation with 12 degrees of freedom \citep{Clayton:2018}. Spatial concordance was assessed between the resampled higher-resolution images (Figure \ref{im1} or \ref{im1gcc}) and the coordinate-registered lower-resolution images (Figure \ref{im2t} or \ref{im2gcct}).

\subsection{Estimating concordance}
For each pair of images, we first calculated Lin's (\citeyear{Lin:1989}) CCC. We then calculated the SCCC as described in Section \ref{sec:local}. We calculated the local concordance coefficient $\rho_i(\cdot)$ in small ($12 \times 12$-pixel) non-overlapping windows. To fit the local model to each small window, we used a Gaussian process $\bm Z(\bm s)=(Z_1(\bm s),Z_2(\bm s))^{\top}, \ \bm s \in \mathbb{R}^2$, with mean $(\mu_1, \mu_2)^{\top}$ and the covariance functions described in equations  \eqref{eq:mat1}--\eqref{eq:Wend}. We used the function \texttt{GeoFit} in the \R \ package \texttt{GeoModels} \citep{GM2018} to compute the ML estimators of the parameters involved in the models. For computational efficiency, the Mat\'ern and Wendland-Gneiting covariances were estimated for a randomly-selected set of ten $20 \times 20$-pixel subimages; the one to be used was selected based on the Akaike and Bayesian Information Critera (AIC and BIC, respectively). In general, the AIC and BIC  coefficients were smaller for estimates of the Mat\'ern covariance than for the Wendland-Gneiting covariance, and so we used the Mat\'ern model even though it took somewhat more time to use it to compute the local estimators. Finally, the global SCCCs for each pair of images were estimated using equations \eqref{eq:rho1} and \eqref{eq:rho2}. 

\subsection{Estimates of concordance}
\begin{figure}[h!]
\begin{center}
\includegraphics[scale=0.45,trim=150 20 150 20,clip]{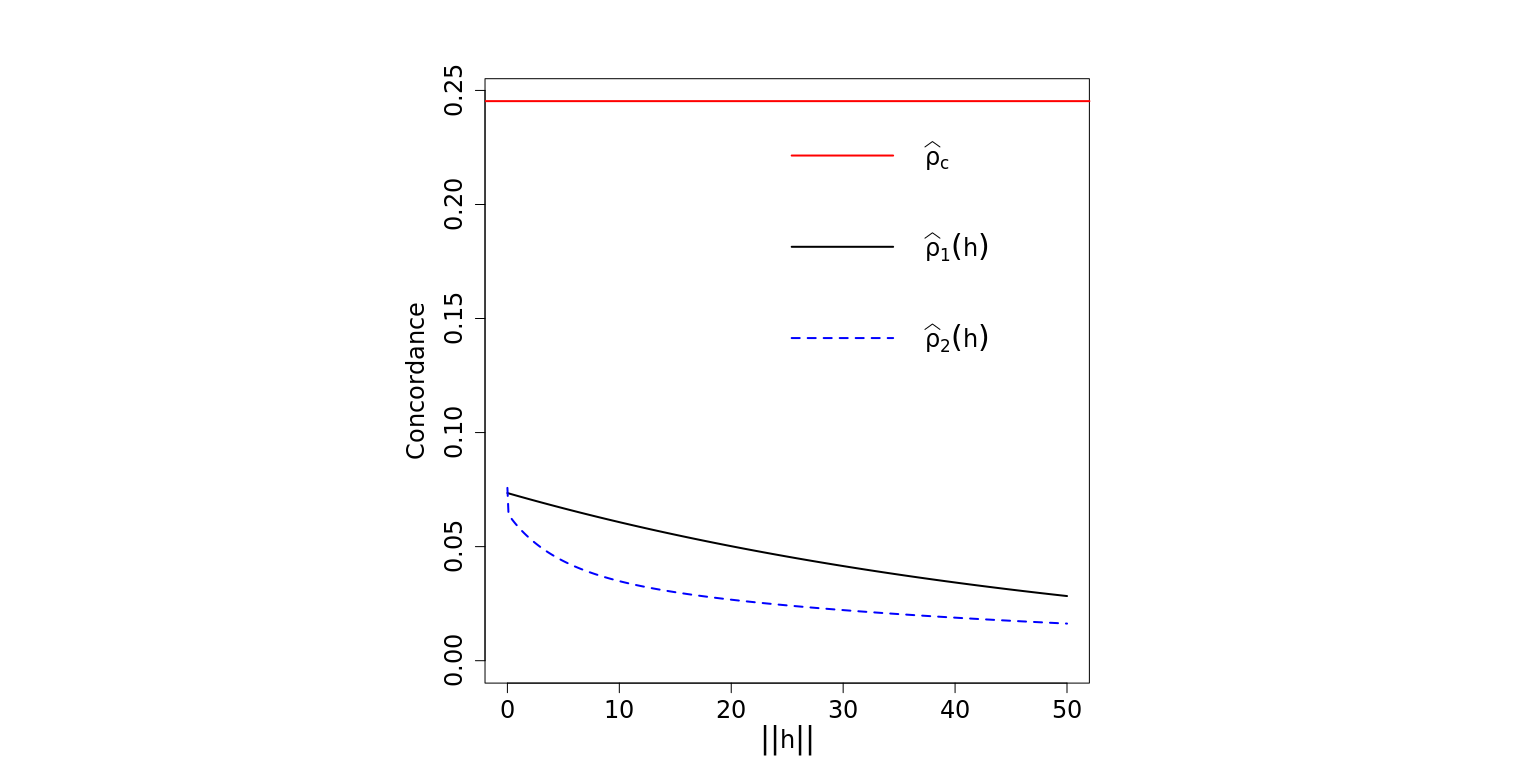}
\end{center}
\vspace{-10pt}
\caption{Global concordance coefficients and Lin's CCC for the \textit{g\textsubscript{cc}}-indexed images.}
\label{cccventanas}
\end{figure}

 Lin's coefficient was $\rho_c=0.1334$ for the grayscale images (Figures \ref{im1} and \ref{im2t}) and $\rho_c=0.2450$ for the \textit{g\textsubscript{cc}}-indexed images  (Figures \ref{im1gcc} and \ref{im2gcct}). In Figure \ref{cccventanas} 
 we plotted Lin's coefficient and the two global coefficients as a function of the spatial norm. We observed a rapid decay of $\widehat{\rho}_2(\cdot)$ and a slower decay of $\widehat{\rho}_1(\cdot)$. The decay was related to the way in which the estimates were computed for each window: $\widehat{\rho}_1(\cdot)$ is a coefficient obtained by plugging in the average of the parameters in the concordance function, but $\widehat{\rho}_2(\cdot)$ is the average of the concordance using all possible windows.

 For $\bm h=\bm 0$, we observed that the SCCC was approximately one-third ($0.08/0.245\times100\approx0.33$) of Lin's CCC. This suggests that Lin's CCC overestimates the spatial concordance between these two images, and implies that would be inappropriate to use it for modeling spatial data.

The images and
all the code used in this paper are available from \url{https://harvardforest.fas.harvard.edu/harvard-forest-data-archive}, dataset HF322.

\section{Discussion}

With the work presented herein, we have extended the standard methodology for estimating concordance into the spatial domain. Our approach consisted in defining a new coefficient that preserves  the interpretation of Lin's (\citeyear{Lin:1989}) concordance correlation coefficient (CCC) for two spatial variables and for a fixed distance lag. Our new spatial concordance correlation coefficient (SCCC) compares the correlation between two spatial variables with respect to their fit to a 45\degree line that passes through the origin. The properties of Lin's (\citeyear{Lin:1989}) CCC are inherited by our SCCC. The ML estimator of our SCCC for the Wendland-Gneiting covariance function is asymptotically normal for an increasing domain sampling scheme. We defined a local SCCC and established its asymptotic normality for the sample version. From the local SCCC, we derived two estimates for the overall SCCC, one based on the average of the \textit{p} local coefficients and the other based on the average of the parameters in the correlation function. Deriving the global SCCC from local coefficients estimated in small non-overlapping windows is computationally more efficient and permits the estimation of spatial concordance for large images that are used commonly in a wide range of applications. 

The Monte Carlo simulation study presented in Section \ref{sec:sim} revealed that for the Mat\'ern and Wendlang-Gneiting covariance functions, the sample version of the SCCC produced accurate estimates of the SCCC that decreased with distance (spatial lag). However, the time required to compute SCCC grows exponentially with window size, implying that for a large image size it is unfeasible to compute $\widehat{\rho}(\cdot)$ using an interpreted language like \R. Although we are exploring ways to improve computational efficiency, the local approach introduced here (Section \ref{sec:local}) appears to be a straightforward way to estimate SCCC for large images.

The camera comparison experiment conducted by \cite{Sonnentag:2012} found that images recorded with a variety of different camera makes and models, all mounted on the top of the same canopy access tower and with a similar field of view, varied in visual appearance, including color balance, saturation, contrast, and brightness. These differences can be attributed to internal differences in sensor design and image processing, and external factors such as lighting. However, \cite{Sonnentag:2012} also found that when simple normalized indices were calculated from the image data, and the emphasis was placed on the seasonality---rather than absolute magnitude---of those indices, the phenological information derived from the imagery was extremely similar across all cameras. Notably, their analysis focused on information about the average color across a large ``region of interest'' drawn across the canopy \citep{Sonnentag:2012}. Although this approach is widely used \citep{Richardson:2019} and it has the advantage of enabling integration across multiple individuals or species that may comprise a typical forest canopy, it lacks spatial information. 

The spatial concordance correlation coefficient we developed and presented here summarizes and accounts for the spatial information in the images, permitting more rigorous characterization of agreement between high-resolution digital images recorded by different sensors. Other applications include using images from different satellite remote-sensing platforms as part of ongoing efforts to harmonize , for example, imagery with different spatial resolution, spectral sensitivity, and angular characteristics \citep[e.g., Landsat-Sentinel efforts][]{Claverie:2018}. Calculation of concordance statistics before and after sensor harmonization could provide critical and objective information about the success of different harmonization methods. There  also could be potential applications in the fusion of remotely-sensed data obtained at different spatiotemporal resolutions, such as MODIS, with its 500-m spatial resolution, daily temporal resolution, and Landsat, with its 30-m spatial resolution, 16-day temporal resolution \citep{Gao:2015}.

\section{Future Work}
Several related theoretical and applied problems arise from the methodology suggested in this article that would be fruitful directions for future research. First, SCCC could be applied to images taken at to points in time by the same camera. The decay of the SCCC as a function of the norm would be expected to be similar to that seen in Figure \ref{cccventanas} for each sequential pair of images. Another approach for dealing with the same problem would be to consider a sequence of \textit{n} images taken with the same camera to be a spatiotemporal process. Then, the SCCC and its estimation properties could be studied in that context. This generalization of the SCCC would have applications in, for example, spatiotemporal analysis of satellite images taken weeks, months, or years apart as a way of characterizing patterns of landscape change. 

\section*{Acknowledgments}

This work has been partially supported by the AC3E, UTFSM, under grant FB-0008, and by grants to A.D.R. from the US National Science Foundation (EF-1065029, EF-1702697, DEB-1237491) and the United States Geological Survey (G10AP00129). This is a contribution from the Harvard Forest Long-term Ecological Research (LTER) site, supported most recently by the US National Science Foundation (DEB 18-32210).

\bigskip
\appendix
\section*{Appendix}
\setcounter{equation}{0}
\renewcommand\theequation{A.\arabic{equation}}

\section*{Mardia and Marshall Theorem}

Let $\left\{Y(\bm{s}):\bm{s}\in D\subset\R^{d}\right\}$ be a Gaussian random field such that $Y(\cdot)$  is observed on $D_n \subset D$.  It is assumed that   $D_n$ is a non-random set satisfying $\|\bm{s}-\bm{t}\|\geq\gamma>0$ for all  $\bm{s},\bm{t}\in D_n$. This ensures that the sampling set is increasing as $n$ increases.
Denote $\bm{Y}=(Y(\bm{s}_1),\dots,Y(\n{s}_n))^\top$ and assume that $\E[\bm{Y}]=\bm{X}\n{\beta},$ $\cov(Y(\bm{t}),Y(\bm{s}))=\sigma(\bm{t},\bm{s};\bm{\theta})$,  $\bm{X}$ is $n\times p$ with  $\text{rank}(\bm X)=p, ~ \n{\beta}\in \mathbb{R}^p$, and $\bm  \theta \in \Theta$, where $\Theta$ is an open set of $\in \mathbb{R}^q$. Let  $\n{\Sigma}=\bm{\Sigma}(\n{\theta})$ be the covariance matrix of $\bm{Y}$ such that  the $ij$-th element of  $\bm{\Sigma}$ is $\sigma_{ij}=\sigma(\bm{s}_i,\bm{s}_j;\bm{\theta})$. We can estimate $\bm \theta$ and $\bm \beta$ using ML, by maximizing
\begin{equation}\label{eq:verosim}
   L=L(\n{\bm \beta},\n{\bm \theta})=k-\dfrac{1}{2}\ln\left|\bm{\Sigma}\right|-\dfrac{1}{2}\left(\n{Y}-\bm{X\beta}\right)^{\top}\bm{\Sigma}^{-1}\left(\bm{Y}-\n{X\beta}\right),
 \end{equation}
where \textit{k} is a constant.

\bigskip

Let
$\bm L_n^{(1)}=\nabla L=(\bm L_{\bm \beta}^{\top},\bm L_{\bm \theta}^{\top})^{\top}$ and
\begin{equation*}
   \bm L_n^{(2)}=\left(\begin{array}{cc} \bm L_{\bm \beta \bm \beta} &\bm  L_{\bm \beta \bm \theta} \\ \bm L_{\bm \theta\bm \beta} & \bm L_{\bm \theta\bm \theta} \end{array}\right)
 \end{equation*}
be the gradient vector and Hessian matrix, respectively, obtained from equation \eqref{eq:verosim}.  Let
$\bm F_n=-\E[\bm L_n^{(2)}]$ be the Fisher information matrix with respect to $\bm \beta$ and $\bm \theta.$ Then,
 $\bm F_n=\diag(\bm F_n{(\bm \beta)},\bm F_n{(\bm \theta})),$ where $\bm F_n{(\bm \beta)}=-\E[\bm L_{\bm \beta\bm \beta}]$ and $\bm F_n{(\bm \theta)}=-\E[\bm L_{\bm \theta \bm \theta}]$.

 For a  twice differentiable covariance function $\sigma(\cdot,\cdot; \bm \theta)$ on $ \Theta$ with continuous second derivatives,
\cite{Mardia:1984} provided sufficient conditions on
 $\bm{\Sigma}$ and $\bm{X}$ such that the limiting distribution of $(\widehat{\bm \beta}_n^{\top},\widehat{\bm \theta}_n^{\top})^{\top}$ is  normal, per the following:
 
\bigskip
\noindent {\bf Theorem.} Let $\lambda_1\leq\cdots\leq\lambda_n$ be the eigenvalues of $\bm{\Sigma}$, and let those of $\bm{\Sigma}_i=\dfrac{\partial\n{\Sigma}}{\partial\theta_i}$ and $\bm{\Sigma}_{ij}=\dfrac{\partial^2\n{\Sigma}}{\partial\theta_i\partial\theta_j}$ be $\lambda_k^{i}$ and $\lambda_k^{ij}$, $k=1,\dots,n,$ such that $|\lambda_1^{i}|\leq\cdots\leq|\lambda_n^{i}|$ and $|\lambda_1^{ij}|\leq\cdots\leq|\lambda_n^{ij}|$ for $i,j=1,\cdots,q$.
Suppose that as $n\rightarrow \infty$
  \begin{enumerate}
     \item[(i)] $\lim\lambda_n=C<\infty$, $\lim|\lambda_n^{i}|=C_i<\infty$ and $\lim|\lambda_n^{ij}|=C_{ij}<\infty$ for all $i,j=1,\dots,q$.
     \item[(ii)] $\|\bm{\Sigma}_{i}\|^{-2}=\mathcal{O}(n^{-\frac{1}{2}-\delta})$ for some $\delta>0$, for $i=1,\dots,q$.
     \item[(iii)] For all $i,j=1,\dots,q$, $a_{ij}=\lim \left[t_{ij}/(t_{ii}t_{jj})^{\frac{1}{2}}\right]$ exists, where $t_{ij}=\Tr\left(\n{\Sigma}^{-1}\bm{\Sigma}_{i}\bm{\Sigma}^{-1}\n{\Sigma}_{j}\right)$ and $\bm{A}=(a_{ij})$ is nonsingular.
     \item[(iv)] $\lim (\bm{X}^\top\bm{X})^{-1}=\bm 0.$
   \end{enumerate}
Then, $(\wh{\bm{\beta}}_n^\top,\wh{\bm{\theta}}_n^\top)^{\top}   \xrightarrow[]{\mathcal{L}} \N\left((\bm{\beta}^\top,\bm{\theta}^\top)^{\top},\bm F_n^{-1}\right)$ as $n\rightarrow \infty$, in an increasing domain sense.

\bigskip

\section*{Proof of Theorem \ref{main}}
The proof consists of verifying the \cite{Mardia:1984} conditions. In Theorem \ref{main},  $\E[{\bm Z(\bm s)}]=\bm 0$; thus the fourth condition in Mardia and Marshall's \citeyear{Mardia:1984} theorem (above), $\lim(\bm X^{\top} \bm X)^{-1}=\bm 0$, is trivially satisfied. Satisfying the first three conditions is somewhat more complex.

For the first two conditions, we start by considering $\nu$ to be fixed. Then
$$\left[ C_{ij} (\bm h,\bm \theta)\right]_{i=j=1}^2 = \left[ \rho_{ij} \sigma_{ii} \sigma_{jj} \left( 1 + (\nu+1) \dfrac{\bm h}{b_{12}} \right) \left( 1 - \dfrac{ \bm h}{b_{12}}\right)_+^{\nu+1} \right]_{i,j=1}^2.$$

Let us consider an increasing domain scenario for  process $\bm Z(\bm s)$, with points  $\bm s_1,...,\bm s_n$ located in a rectangle  $D_n \subset \Delta \mathbb{Z}^d$, for $0<\Delta<L$, and $D_n \subset D_{n+1}$, for all $n$.

 Define the distance matrix   $\bm H_n = \left[H_{lq} \right]_{l=q=1}^n$, where $ H_{lq}= \| \bm s_l-\bm s_q \|$, and $\|\cdot \|$ denotes the Euclidean norm. Then the covariance matrix of $(\bm Z(\bm s_1)^{\top}, \ldots, \bm Z(\bm s_n)^{\top})^{\top}$ can be written as
$$\bm \Sigma_n (\bm \theta)= \begin{pmatrix}
\sigma_1^2 & \sigma_1 \sigma_2 \rho_{12}\\ - & \sigma_2^2
\end{pmatrix} \otimes \bm \Gamma_n,
$$
where $\bm \Gamma_n = \left[ \left(1+ \dfrac{(\nu+1) H_{lq}}{b_{12}} \right) \left(1- \dfrac{ H_{lq}}{b_{12}} \right)_+^{\nu+1} \right]_{l=q=1}^n$ \ and $\bm \theta = (\sigma_1^2,\sigma_2^2,\rho_{12},b_{12})^\top$. Taking derivatives, we obtain
\begin{align*}
\dfrac{\partial \bm \Sigma_n (\bm \theta)}{\partial \sigma_1^2} & = \begin{pmatrix}
1 & \dfrac{\sigma_2 \rho_{12}}{2 \sigma_1}\\
- & 0
\end{pmatrix} \otimes \bm \Gamma_n,
  & \dfrac{\partial \bm \Sigma_n (\bm \theta)}{\partial \sigma_2^2} &=  \begin{pmatrix}
0 & \dfrac{\sigma_1 \rho_{12}}{2 \sigma_2} \\
- & 1
\end{pmatrix} \otimes \bm \Gamma_n, \\
\dfrac{\partial \bm \Sigma_n (\bm \theta)}{\partial b_{12}} & = \begin{pmatrix}
\sigma_1^2 & \sigma_1 \sigma_2 \rho_{12} \\
- & \sigma_2^2
\end{pmatrix} \otimes \bm S_n,  & \dfrac{\partial \bm \Sigma_n (\bm \theta)}{\partial \rho_{12}} &=\begin{pmatrix}
0 & \sigma_1 \sigma_2\\ - & 0
\end{pmatrix} \otimes \bm \Gamma_n,
\end{align*}
where $\bm S_n$ is given by $$\bm S_n=\dfrac{\partial \bm \Gamma_n}{\partial b_{12}}= \left[ \dfrac{(\nu+1) H_{lq}}{b_{12}^2} \left( 1- \dfrac{H_{lq}}{b_{12}} \right)_+^\nu \left( - \left( 1- \dfrac{H_{lq}}{b_{12}}\right)_+ + \left(1+ \dfrac{(\nu+1) H_{lq}}{b_{12}} \right) \right) \right]_{l=q=1}^n.$$

For any matrix norm, the spectral radius 
 $\lambda_{\text{max}} \lbrace \bm A \rbrace$ of an $n \times n$ matrix $\bm A$ satisfies
  $\lambda_{\text{max}} \lbrace \bm A \rbrace \leq \| \bm A \|$. Then, considering the norm  $\| \cdot \|_{\infty}$, we have
\begin{align*}
\lambda_{\text{max}} \lbrace \bm \Gamma_n \rbrace \leq \| \bm \Gamma_n \|_{\infty} &= \max_l \sum_{q=1}^n \left| \left( 1+ \dfrac{(\nu+1) H_{lq}}{b_{12}} \right) \left( 1- \dfrac{H_{lq}}{b_{12}} \right)_+^{\nu+1} \right|\\
&= \sup_{1\leq l \leq n} \sum_{q=1}^n \left| \left( 1+ \dfrac{(\nu+1) H_{lq}}{b_{12}} \right) \left( 1- \dfrac{H_{lq}}{b_{12}} \right)_+^{\nu+1} \right| \\
&< \sum_{s \in \Delta \mathbb{Z}^d}  \left( 1+ \dfrac{(\nu+1) \| s \|}{b_{12}} \right)  \left( 1- \dfrac{\|\bm s \|}{b_{12}} \right)_+^{\nu+1} .
\end{align*}
One can check that $$\int_{s \in \mathbb{R}^d} \left( 1+ \dfrac{(\nu+1) \| s \|}{b_{12}} \right)  \left( 1- \dfrac{\|\bm s \|}{b_{12}} \right)_+^{\nu+1} ds < \infty.$$ 
Thus $\sup_n \lambda_{\text{max}} \lbrace \bm \Gamma_n \rbrace<\infty$, which implies that  $\sup_n \lambda_{\text{max}} \lbrace \bm \Sigma_n (\bm \theta) \rbrace<\infty$. Because $\bm \Gamma_n$ is positive definite, $\lambda_i \lbrace \bm \Gamma_n \rbrace > 0$, $i=1,...,n$. In particular, $\lambda_{\text{min}} \lbrace \bm \Gamma_n \rbrace >0$, so $\inf_n \lambda_{\text{min}}  \lbrace \bm \Gamma_n \rbrace >0$ and $\inf_n \lambda_{\text{min}} \lbrace \bm \Sigma_n \rbrace >0$. Further, $$ \sup_n   \lambda_{\text{max}} \left\{ \dfrac{\partial \bm \Sigma_n (\bm \theta)}{\partial \sigma_1^2} \right\} = \sup_n   \lambda_{\text{max}} \left[ \begin{pmatrix}
1 & \dfrac{\sigma_2 \rho_{12}}{2 \sigma_1} \\
- & 0
\end{pmatrix} \otimes \bm \Gamma_n
 \right] < \infty, 
\ \text{for} \ \dfrac{\sigma_2 \rho_{12}}{2 \sigma_1} < \infty.$$

Similarly,  $$\sup_n   \lambda_{\text{max}} \left\{ \dfrac{\partial \bm \Sigma_n (\bm \theta)}{\partial \sigma_2^2} \right\} ,\sup_n   \lambda_{\text{max}} \left\{ \dfrac{\partial \bm \Sigma_n (\bm \theta)}{\partial \rho_{12}} \right\} < \infty.$$

Moreover, $\lambda_{\text{max}} \lbrace \bm S_n \rbrace \leq \| \bm S_n \|_{\infty}< \infty$ because of the form of the polynomial in  $\bm s \in \mathbb{R}^d$ and the compact support in $b_{12}$. Then, for $\sigma_1^2, \sigma_2^2, \sigma_1 \sigma_2 \rho_{12} < \infty$,  $$\sup_n   \lambda_{\text{max}} \left\{ \dfrac{\partial \bm \Sigma_n (\bm \theta)}{\partial b_{12}} \right\}  < \infty.$$

This implies that,  $$\sup_n   \lambda_{\text{max}} \left\{ \dfrac{\partial \bm \Sigma_n (\bm \theta)}{\partial \theta_c} \right\}  < \infty, \quad c=1,2,3,4.$$
 
The second derivatives are:
\begin{align*}
\dfrac{\partial^2 \bm \Sigma_n (\bm \theta)}{\partial \sigma_1^2 \partial \sigma_2^2} & = \begin{pmatrix}
0 & \dfrac{\rho_{12}}{4 \sigma_1 \sigma_2}\\
- & 0
\end{pmatrix} \otimes \bm \Gamma_n,
  & \dfrac{\partial \bm \Sigma_n (\bm \theta)}{\partial \sigma_1^2 \partial b_{12}} &=  \begin{pmatrix}
1 & \dfrac{\sigma_2 \rho_{12}}{2 \sigma_1} \\
- & 0
\end{pmatrix} \otimes \bm S_n, \\
\dfrac{\partial \bm \Sigma_n (\bm \theta)}{\partial \sigma_1^2 \partial \rho_{12}} & = \begin{pmatrix}
0 & \dfrac{\sigma_2}{2 \sigma_1}  \\
- & 0
\end{pmatrix} \otimes \bm \Gamma_n,  & \dfrac{\partial \bm \Sigma_n (\bm \theta)}{\partial \sigma_1^4} &=\begin{pmatrix}
0 & - \dfrac{\sigma_2 \rho_{12}}{4 \sigma_1^3} \\ - & 0
\end{pmatrix} \otimes \bm \Gamma_n,\\
\dfrac{\partial \bm \Sigma_n (\bm \theta)}{\partial \sigma_2^2 \partial b_{12}} & = \begin{pmatrix}
0 & \dfrac{\sigma_1 \rho_{12}}{2 \sigma_2}\\
- & 1
\end{pmatrix} \otimes \bm S_n,
  & \dfrac{\partial \bm \Sigma_n (\bm \theta)}{\partial \sigma_2^2 \partial \rho_{12}} &=  \begin{pmatrix}
0 & \dfrac{\sigma_1 }{2 \sigma_2} \\
- & 0
\end{pmatrix} \otimes \bm \Gamma_n, \\
\dfrac{\partial \bm \Sigma_n (\bm \theta)}{\partial \sigma_2^4} & = \begin{pmatrix}
0 & -\dfrac{\sigma_1 \rho_{12}}{4 \sigma_2^3} \\
- & 0
\end{pmatrix} \otimes \bm \Gamma_n,  & \dfrac{\partial \bm \Sigma_n (\bm \theta)}{\partial b_{12} \partial \rho_{12}} &=\begin{pmatrix}
0 & \sigma_1 \sigma_2\\ - & 0
\end{pmatrix} \otimes \bm S_n,\\
\dfrac{\partial \bm \Sigma_n (\bm \theta)}{\partial b_{12}^2} & = \begin{pmatrix}
\sigma_1^2 & \sigma_1 \sigma_2 \rho_{12} \\
- & \sigma_2^2
\end{pmatrix} \otimes \bm {SS}_n,  & \dfrac{\partial \bm \Sigma_n (\bm \theta)}{\partial \rho_{12}^2} &=0,
\end{align*}
where $\bm {SS}_n = \dfrac{\partial}{\partial b_{12}} \bm S_n$.

Because $\sup_n \lambda_{\text{max}} \lbrace \bm 0 \rbrace < \infty$, the compact support of  $\bm {SS}_n$ in $b_{12}$, and the previous results,  $\lambda_{\text{max}}  \lbrace  \bm  {SS}_n \rbrace  \leq  \|\bm {SS}_n \|_{\infty}  <  \infty$. Then, for $\sigma_1^2, \sigma_2^2, \sigma_1 \sigma_2 < \infty$, 
$$\sup_n \lambda_{\text{max}} \left\{ \dfrac{\partial^2 \bm \Sigma_n(\bm \theta)}{\partial b_{12}^2} \right\} < \infty.$$
 In addition, $$\left\| \dfrac{\partial \bm \Sigma_n(\bm \theta)}{\partial \theta_i } \right\|_\infty \leq \left\| \dfrac{\partial \bm \Sigma_n(\bm \theta)}{\partial \theta_i } \right\| \leq \sqrt{n} \left\| \dfrac{\partial \bm \Sigma_n(\bm \theta)}{\partial \theta_i } \right\|_\infty.$$
 
This satisfies the first two conditions of Mardia and Marshall's theorem.

For the third condition, we consider $\bm A= [a_{ij}]_{i=j=1}^p$, with $a_{ij} = \left\{ \dfrac{t_{ij}}{(t_{mm}t_{nn})^{1/2}} \right\},$ and $t_{ij} = \text{tr} \left\{ \bm \Sigma_n (\bm \theta)^{-1} \dfrac{\partial \bm \Sigma_n (\bm \theta)}{\partial \theta_i}  \bm \Sigma_n (\bm \theta)^{-1} \dfrac{\partial \bm \Sigma_n (\bm \theta)}{\partial \theta_j}  \right\} \ \text{for all} \ i,j=1,...,p$; we prove that  $\bm A$ is non singular.

Notice that

$$ \bm T = [t_{ij}]_{i=j=1}^4 = \begin{pmatrix}
\dfrac{n(\rho_{12}^2-2)}{4 \sigma_1^4(\rho_{12}^2-1)}& \dfrac{n \rho_{12}^2}{4 \sigma_1^2 \sigma_2^2(\rho_{12}^2-1)} &\dfrac{1}{2 \sigma_1^2} \text{tr} \lbrace \bm A_n \rbrace & \dfrac{n \rho_{12}}{2 \sigma_1^2 (\rho_{12}^2-1)} \\
-& \dfrac{n(\rho_{12}^2-2)}{4 \sigma_2^4(\rho_{12}^2-1)}& \dfrac{1}{2 \sigma_2^2} \text{tr} \lbrace \bm A_n \rbrace &  \dfrac{n \rho_{12}}{2 \sigma_2^2 (\rho_{12}^2-1)} \\
-&- & \text{tr} \lbrace [\bm A_n]^2 \rbrace & \dfrac{\rho_{12}}{\rho_{12}^2-1} \text{tr} \lbrace \bm A_n \rbrace  \\
-&- &- & \dfrac{n (\rho_{12}^2+1)}{(\rho_{12}^2-1)^2}
\end{pmatrix},
$$
with $\bm A_n= \lbrace \bm \Gamma_n^{-1} \circ \bm  S_n \rbrace$ where the operator $\circ$ denotes the matrix Hadamard product.

Then, 

\begin{equation}\label{eq:A} \bm A = \begin{pmatrix}
1& \dfrac{\rho_{12}^2}{\rho_{12}^2-2} & \dfrac{\text{tr}(\bm A_n)}{\left( \dfrac{n(\rho_{12}^2-2) \text{tr}([\bm A_n]^2)}{\rho_{12}^2-1} \right)^{1/2}} & \dfrac{\rho_{12}}{\left( \dfrac{(\rho_{12}^2-2)(\rho_{12}^2+1)}{\rho_{12}^2-1} \right)^{1/2}} \\
-&1 &  \dfrac{\text{tr}(\bm A_n)}{\left( \dfrac{n(\rho_{12}^2-2) \text{tr}([\bm A_n]^2)}{\rho_{12}^2-1} \right)^{1/2}} & \dfrac{\rho_{12}}{\left( \dfrac{(\rho_{12}^2-2)(\rho_{12}^2+1)}{\rho_{12}^2-1} \right)^{1/2}}  \\
-&- &1 & \dfrac{-\rho_{12} \text{tr}(\bm A_n)}{(n \text{tr}([\bm A_n]^2)(\rho_{12}^2+1))^{1/2}}\\
-&- &- &1
\end{pmatrix}.
\end{equation}

For matrix $\bm A$ in equation \eqref{eq:A},we have extended the result established by  \cite{Bevilacqua:2015}. Thus  $\bm A$ is positive definite. By Mardia and Marshall's Theorem the ML estimator  of   $\bm \theta = (\sigma_1^2,\sigma_2^2,\rho_{12},b_{12})^\top$ is asymptotically normal with variance  $\bm F_n (\bm \theta)^{-1}$.

Equation \eqref{aux1} implies that

$$\rho^c (\bm h) = g(\bm \theta)=\dfrac{2 \rho_{12} \sigma_1 \sigma_2 \left( 1 + (\nu +1) \dfrac{\| \bm h\|}{b_{12}} \right) \left( 1 - \dfrac{\| \bm h\|}{b_{12}}\right)_+^{\nu+1}  }{\sigma_1^2 + \sigma_2^2}.$$ 
Fixing $\nu>0$, noting that $g(\cdot)$ is a continuously differentiable function for 
$\sigma_1\neq$ 0 and $\sigma_2\neq 0$,
and using the multivariate delta method for $g(\cdot)$  we obtain
$$
\left( \nabla g(\bm \theta)^\top \bm F_n ( \bm \theta)^{-1} \nabla g(\bm \theta) \right)^{-1/2} (g(\bm \theta_n) - g(\bm \theta)) \xrightarrow[]{D} N(0, 1 ),
$$
where $$\nabla g(\bm \theta) = \begin{pmatrix}
\dfrac{\sigma_2 \rho_{12}(\sigma_2^2-\sigma_1^2) \left( 1 + (\nu +1) \dfrac{\| \bm h\|}{b_{12}} \right) \left( 1 - \dfrac{\| \bm h\|}{b_{12}}\right)_+^{\nu+1}}{\sigma_1 (\sigma_1^2+\sigma_2^2)^2}\\
\dfrac{\sigma_1 \rho_{12}(\sigma_1^2-\sigma_2^2) \left( 1 + (\nu +1) \dfrac{\| \bm h\|}{b_{12}} \right) \left( 1 - \dfrac{\| \bm h\|}{b_{12}}\right)_+^{\nu+1}}{\sigma_2 (\sigma_1^2+\sigma_2^2)^2}\\
\dfrac{2 \sigma_1 \sigma_2 \left( 1 + (\nu +1) \dfrac{\| \bm h\|}{b_{12}} \right) \left( 1 - \dfrac{\| \bm h\|}{b_{12}}\right)_+^{\nu+1}}{\sigma_1^2+\sigma_2^2}\\
\dfrac{2 \sigma_1 \sigma_2 \rho_{12} f(b_{12})}{\sigma_1^2+\sigma_2^2}
\end{pmatrix},$$
and
{\small $f(b_{12})=  \left( -\dfrac{(\nu+1) \| \bm h \|}{b_{12}^2} \right) \left( 1- \dfrac{\| \bm h\|}{b_{12}} \right)^{\nu+1}_+ + \left( 1+  \dfrac{ (\nu+1)\|\bm h\|}{b_{12}}\right) \left(1-\dfrac{\|\bm h\|}{b_{12}} \right)^\nu_+ \dfrac{(\nu+1) \| \bm h\|}{b_{12}^2}  .$} \hfill $\square$

\section*{References}

\bibliography{main.bib}

\end{document}